\newcounter{myctr}

\documentclass{ETHpaper}
\usepackage{url}
\usepackage{graphicx, epsfig, amsmath, amssymb,wasysym}
\usepackage[pdftex,dvipsnames]{xcolor}
\usepackage[square,numbers,sort&compress]{natbib}
\usepackage[justification=justified,labelfont=bf]{caption}
\usepackage{soul}
\usepackage{cancel}
\usepackage{xargs}  
\usepackage[colorinlistoftodos,prependcaption,textsize=tiny]{todonotes}
\newcommandx{\unsure}[2][1=]{\todo[linecolor=red,backgroundcolor=red!25,bordercolor=red,#1]{#2}}
\newcommandx{\change}[2][1=]{\todo[linecolor=blue,backgroundcolor=blue!25,bordercolor=blue,#1]{#2}}
\newcommandx{\improvement}[2][1=]{\todo[linecolor=OliveGreen,backgroundcolor=OliveGreen!25,bordercolor=OliveGreen,#1]{#2}}
\newcommandx{\info}[2][1=]{\todo[linecolor=Plum,backgroundcolor=Plum!25,bordercolor=Plum,#1]{#2}}
\newcommandx{\thiswillnotshow}[2][1=]{\todo[disable,#1]{#2}}

\begin{document}

\newcommand{\mean}[1]{\left\langle #1 \right\rangle} \newcommand{\abs}[1]{\left| #1 \right|}
\renewcommand{\epsilon}{\varepsilon} \newcommand{\eps}{\varepsilon} \renewcommand*{\=}{{\kern0.1em=\kern0.1em}}
\renewcommand*{\-}{{\kern0.1em-\kern0.1em}} \newcommand*{\+}{{\kern0.1em+\kern0.1em}}

\renewcommand{\thefootnote}{ \fnsymbol{footnote} }

\title{How do OSS projects change in number and size? \\ 
A large-scale analysis to test a model of project growth}

\titlealternative{How do OSS communities change in number and size?}

\author{Frank Schweitzer\footnote{Corresponding author: \url{fschweitzer@ethz.ch}}, Vahan Nanumyan, Claudio J. Tessone
, Xi Xia}

\authoralternative{F. Schweitzer, V. Nanumyan, C.J.~Tessone, X. Xia}

\address{Chair of Systems Design, ETH Zurich, Weinbergstrasse 58, 8092 Zurich, Switzerland}

 \reference{Published in \href{http://www.worldscientific.com/doi/abs/10.1142/S0219525915500083}{\emph{Advances in Complex Systems}, 17, 1550008 (2014)}.} 


\makeframing

\maketitle

\begin{abstract}
Established Open Source Software (OSS) projects can grow in size if new developers join, but also the number of OSS projects can grow if developers choose to found new projects. 
We discuss to what extent an established model for firm growth can be applied to the dynamics of OSS projects. 
Our analysis is based on a large-scale data set from SourceForge (SF) consisting of monthly data for 10 years, for up to 360\,000 OSS projects and up to 340\,000 developers. 
Over this time period, we find an exponential growth both in the number of projects and developers, with a remarkable increase of single-developer projects after 2009. 
We analyze the monthly entry and exit rates for both projects and developers, the growth rate of established projects and the monthly project size distribution. 
To derive a prediction for the latter, we use modeling assumptions of how newly entering developers choose to either found a new project or to join existing ones. 
Our model applies only to collaborative projects that are deemed to grow in size by attracting new developers. 
We  verify, by a thorough statistical analysis, that the Yule-Simon distribution is a valid candidate for the size distribution of collaborative projects except for certain time periods where the modeling assumptions no longer hold. 
We detect and empirically test  the reason for this limitation, i.e., the fact that an increasing number of established developers found additional new projects after 2009. 
\end{abstract}

\date{\today}

\section{Introduction}
\label{sec:Introduction}

\emph{Open Source Software} (OSS) communities share with other organizations, such as social online platforms \citep{Szell2010} or research and development networks \citep{Schweitzer2014} the feature that they are inherently \emph{dynamic} because of the continuous \emph{entry} of new members (developers, users, firms) and \emph{exit} of established members. 
While this entry and exit dynamics usually resembles small perturbations that do not challenge the existence of the organization, it can also lead to large cascades of members leaving \citep{Mavrodiev2013}, in particular if these depend on the contribution of those who left. 
Hence, these processes have the potential to destabilize an organization. 
On the other hand, the entry-exit dynamics plays an important role in \emph{knowledge exchange} between organizations. 
New members can bring new knowledge, information, skills, or methods that help organizations to innovate. 
Members leaving, on the other hand, make space for newcomers and at the same time transfer knowledge they gained to other organizations. 

The economist Schumpeter saw the \emph{creative destruction} process induced by newcomers as an important element to renew, and to develop, the economic system \citep{schumpeter1911}.
Consequently, economists have for a long time focused on the role of entry and exit of \emph{firms} in industrial organization \citep{mansfield_1962}. 
For example, they found positive correlations between the entry rate of firms and innovation rates \citep{geroski1995we}. 
A particular strand of research was devoted to the impact of newcomers on the \emph{size distribution} of firms. 
This is a long-standing topic in industrial organization since Gibrat \citep{sutton} introduced the \emph{law of proportionate growth}, i.e., $\dot{x}=\beta\,x$, where $x$ represents the firm size as measured by the number of employees, to explain the empirical size distribution of firms. 
His assumptions lead to a \emph{log-normal distribution} which is valid only if the number of firms is kept constant.
An important extension was made by Simon \citep{simon1955} who combined the model of proportionate growth with assumptions about firm's entry.
This yields another type of skew size distribution, which he called the \emph{Yule} distribution \citep{yule1925mathematical}, but is now commonly called the \emph{Yule-Simon distribution}. 
It is characterized by a power-law tail, $f(x)\propto x^{-\gamma}$, for large values of $x$. 

The debate about whether the firm size distribution is best described by a log-normal, Yule-Simon, or a power-law distribution is still ongoing and the answer largely depends on the dataset analyzed.  
Therefore, we focus more on the theoretical insights obtained from these investigations. 
In particular, we ask to what extent an \emph{economic model}, i.e., the Simon model for the entry and subsequent growth of \emph{firms}, can be utilized to describe the dynamics in other types of social organizations, for example \emph{OSS communities}.

It would indeed add to the importance of the Simon model if we find that it also describes the empirical findings in the dynamics of OSS communities. 
On the other hand, a formal model of the entry dynamics and growth of OSS communities which focuses on the choice of developers is a rather new and important contribution to better understand the complex processes in socio-technical systems \citep{Tessone2010, geipel-acs}.
Precisely, the novel contribution of our paper is \emph{not} in the development of the model, but in the discussion to what extent an existing economic model describes the dynamics in OSS communities and how it could be extended for this purpose. 

The methodological approach to test a model of firm growth for OSS communities is based on some implicit analogies. With OSS \emph{community}, we refer to a specific platform that hosts possibly hundreds of thousands of OSS projects, such as \texttt{Sourceforge.net} or \texttt{Github.com}. 
I.e., the community is comprised of \emph{projects} each of which has a number of \emph{developers} contributing. 
We note that such as system is best described as a bipartite network as discussed in Sect. \ref{sec:bipartite}. 
Continuing with the analogy to industrial organization, this system is equivalent to  a particular \emph{industrial sector} (also called market). 
This market is comprised of thousands of \emph{firms} each of which has a number of \emph{employees}. The \emph{size} of the firm is given by the number of employees, as the size of the project is given by the number of developers. 

With respect to the dynamics, we observe a continuous entry of new firms/projects that into the market/platform, but likewise also a continuous exit, e.g., if firms go bankrupt or projects collapse. 
But it is not the firms/projects drive the dynamics. The real drivers are the underlying constituting elements, i.e., the employees/developers that create new firms/projects or join existing ones, or decide to quit. 
This leaves a considerable degree of freedom. 
Employees/developers usually decide individually which firm/project to join or whether to establish a \emph{new} firm/project. 
Only the latter choice leads to an increase in the total number of firms/projects, while the former still results in an increase in the size of a given firm/project.
Interestingly, on the system's level this individual choice can be described by a certain probability to found a new firm/project, which is constant and the same across the population.
While this does not reflect the individual motivation, it is sufficient to describe the dynamics on the system's level. 

Hence, with a focus on the \emph{OSS community} we are not so much interested in the individual dynamics of specific projects which would be better captured in case studies \citep{Scholtes2013}.
Instead, we want to investigate systemic properties that result from a large number of projects. 
Such an approach does not necessarily address a number of issues that may be also of interest in the study of OSS communities, such as the motivation of developers \citep{vkrogh-2015}, their individual activity \citep{github} or their specific role/skills in the project. 

Our paper is organized as follows: 
Before we propose in, Sect. \ref{sec:macro}, a model to capture the dynamics of projects (and indirectly also of developers), in Sect. \ref{sec:analysis} we  look at the macroscopic properties of the community, obtained by aggregating over all projects. 
In Sect. \ref{sec:micro} we also analyze in detail the entry and exit rates of projects and developers and particularly focus on the size dependence of growth rates. 
The model we develop leads to a prediction for the size distribution of projects, which is compared with empirical data in Sect. \ref{sec:disc}. There, we also discuss reasons for deviations from the prediction and possible extensions of our work.

\section{Aggregated data analysis}
\label{sec:analysis}

\subsection{Dataset description}
\label{sec:data}

The dataset used in our study was acquired from \texttt{Sourceforge.net} (SF), which was one of the world's largest Open Source software development website until \texttt{Github.com} became predominant. 
Our analyzed dataset contains 89 monthly snapshots from January 2003 to June 2012, in which information about  all the developers and projects hosted on SF is recorded.
In their early years, SF frequently changed the format in which information about the project was stored. 
This  leads to disruptions in our dataset because of corrupt data, in particular snapshots between February 2003 and October 2004 and  for January 2005 are missing. 
Also, snapshots for July, August and September 2007 were removed by the SF archive provider because of data corruption \cite{Reference7}. 
Eventually, in February 2006, SF launched an Autopurge Service to remove inactive projects, which resulted in abnormal dropdowns in the number of projects.
Starting from June 2010, SF automatically created a project for each developer.
These ``projects'' do not represent real activities of the developers, so we removed them from the analysis.
Also, in total 3 developers/projects were manually removed from the dataset.
These three points had an extremely high number of links, were created by machines, and were used for the purpose of advertising or testing.

Nevertheless, the remaining dataset is large and reliable enough for our analysis. 
Table \ref{table:variables} gives an overview of the total number of projects, $N_{p}(t)$, and developers, $N_{d}(t)$, in the first and the last month recorded in our dataset. 
We also have information about the relationship between projects and developers, in particular about the \emph{entry date} (month) in which a developer joined a project. 
We then assume that a \emph{link} between the developer and the project was created.
The total number of links between developers and projects, $K(t)$, is also reported in Table \ref{table:variables}.
We note that, due to the lack of data, the programming language used is available only for about 40\% of all projects.
\begin{table}[hbt]
  \centering
  \caption{Summary of available data for the first and the last monthly snapshot of the dataset.
$N_{d}(t)$: \emph{total number of developers} at time $t$, $N_{p}(t)$; \emph{total number of projects} at time $t$, $K(t)$: \emph{total number of links} between developers and projects at time $t$}
  \label{table:variables}
  \begin{tabular}{|c|r|r|}\hline
    Time & 01/2003 & 07/2012\\
    \hline \hline
    $N_{d}(t)$ & 77.050 & 339.140 \\ \hline
    $N_{p}(t)$ & 54.234 & 357.555 \\ \hline
$K(t)$ & 106.840 & 576.238\\
    \hline
  \end{tabular} 
\end{table}

\subsection{Aggregated growth rates}
\label{sec:growth}

\begin{figure}[htb]
  \centering
  \includegraphics[width=0.49\textwidth]{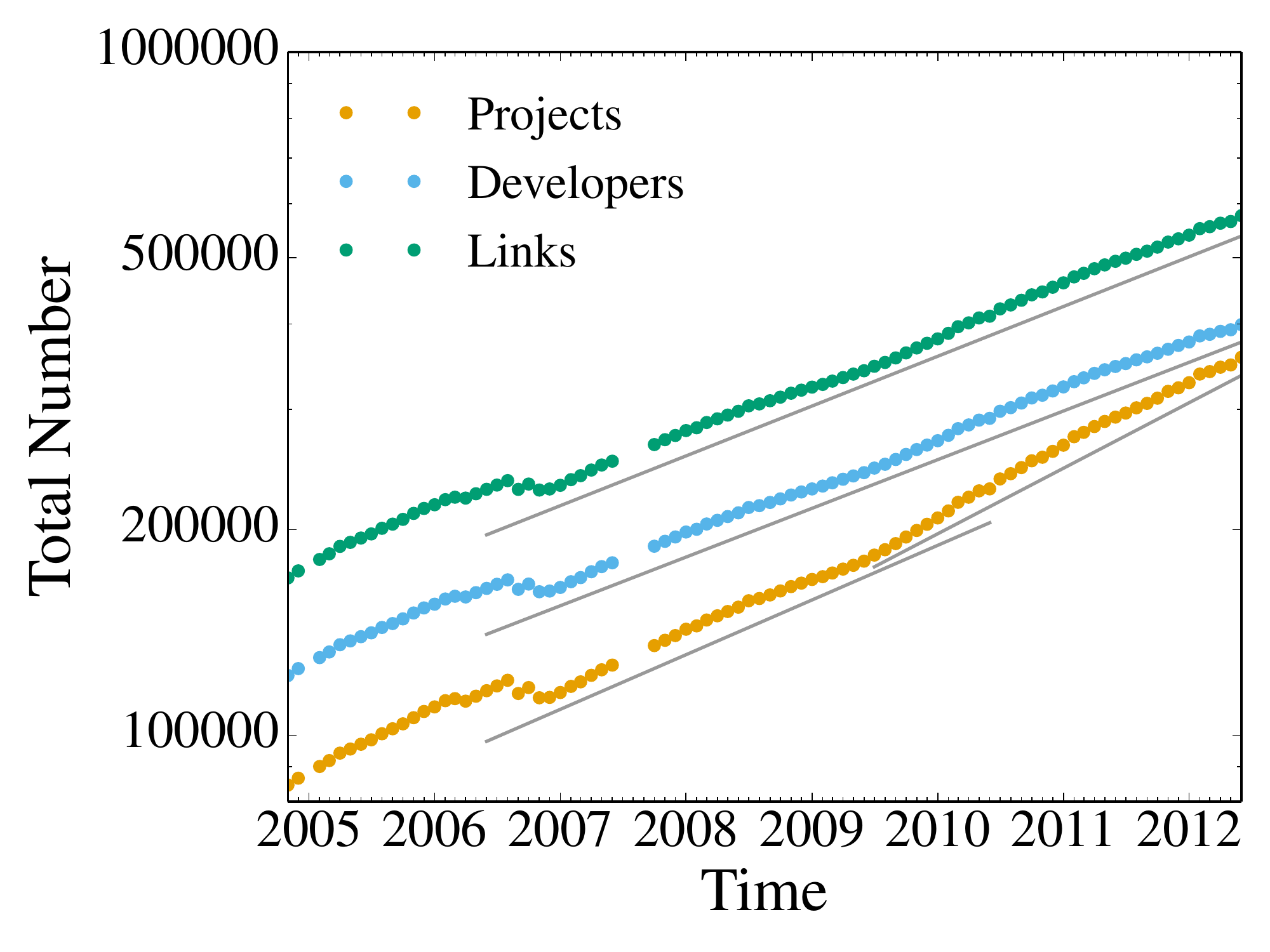}\hfill
\includegraphics[width=0.49\textwidth,page=2]{new_collab_vs_all}
  \caption{
\textbf{(left)} Total number of projects, $N_{p}(t)$ (yellow), total number of developers, $N_{d}(t)$ (blue), and total number of links, $K(t)$ (green), over time measured in months. The solid lines indicate fits of the growth rates given in Table \ref{number regression}. \textbf{(right)} Number of new single-developer projects (yellow) and multi-developer projects (blue) per month, over time. Solid lines represent the median obtained over a rolling window of one year. The EM estimate (green) is discussed in Sect. \ref{sec:rhop0}.
 The missing points in 2006 correspond to the time periods when Autopurge was used excessively (see Sect. \ref{sec:data}).
}
  \label{fig:OSS Growth Vs. Time}
\end{figure}

The most simple aggregated statistics is given by the total number of projects, $N_{p}(t)$, developers, $N_{d}(t)$ and links, $K(t)$, and how these numbers change over time measured in \emph{months}.
Figure~\ref{fig:OSS Growth Vs. Time}(left) shows their evolution.
As we clearly observe in the log-linear plot, all of these quantities follow an \emph{exponential growth} dynamics:
\begin{equation}
  \label{eq:exp}
  \frac{\Delta X}{\Delta t}=\omega X\;; \quad X(t) \propto \exp{\left\{\omega t\right\}}
\end{equation}
This is also known as the \emph{law of proportionate growth} and indicates that the SF community became more attractive the larger it was, which reamplified the growth for many years.
The respective \emph{growth rates} $\omega$ with $\Delta t$=1 month are given in Table \ref{number regression}.

\begin{table}[htb]
  \centering
  \caption{Regression results for Fig.~\ref{fig:OSS Growth Vs. Time}(left)}
  \label{number regression}
  \begin{tabular}{|c|c|c|c|}
    \hline
    Variable & Growth Rate $\omega$ & $R^2$ & p-value \\ \hline \hline
    $K(t)$ & 1.30\% & $>0.99$ & 2.80e-99 \\ \hline
    $N_{d}(t)$ & 1.27\% & $>0.99$ & 8.18e-99 \\ \hline
    $N_{p}(t)$ & 1.54\% &$>0.99$ & 8.06e-82 \\ \hline \hline
    $N_{p}(t<2010)$ & 1.33\% &$>0.99$ &6.19e-55 \\ \hline
    $N_{p}(t>2010)$ & 1.81\% & $>0.99$ & 2.30e-33 \\ \hline 
  \end{tabular} 
\end{table} 

We note that the exponential growth remains despite of the data disruptions explained in  Section \ref{sec:data}.
A closer look at Fig.~\ref{fig:OSS Growth Vs. Time} and Table \ref{number regression} reveals that, in the log-linear plot, both $N_{d}(t)$ and $K(t)$ grow at about the same growth rate, constant over time.
For $N_{p}(t)$, however, we observe a significant change in the growth rate at about 2010.
Before 2010, $N_{p}(t)$ grew at a rate comparable to the other aggregated quantities, but the growth significantly increased afterwards.
Remarkably, this increase does not become visible in the growth of the total number of links.
Hence, the network between developers and projects (to which the links refer) becomes sparser after 2010.

We argue that the increasing growth rate for projects results from the fact that developers started their own single-developer  projects. 
These could be either new developers entering SF or developers who already registered at SF for another project and now create another one. 
This conjecture is explored in Fig. \ref{fig:OSS Growth Vs. Time}(right) which plots the number of new projects per month that have only one developer together with the respective number of new projects per month that have more than one developer. 
We observe a significant increase of single-developer projects at about 2010, while the number of new multi-developer projects per month remain about the same over time. 

\subsection{Change in programming languages}
\label{sec:lang}

One of the reasons for the observed change towards more single-developer projects could be in the rise of \emph{scripting languages} for programming, such as \texttt{PHP} and, more recently, \texttt{Python}.
Such programming languages have been widely adopted in particular for single developer projects, as we verify in our dataset. 
We already mentioned that only about 40\% of all projects list their programming language and some projects, especially large ones, also use more than one programming language. 
Precisely, in 01/2003 information about the programming language was available for 35.089 projects, which increased to 187.168 projects in 07/2012.
There are in total 106 programming language listed in the dataset, but more than 80\% of all projects use one of the major 7 languages \texttt{C}, \texttt{C\#}, \texttt{C++}, \texttt{Python}, \texttt{PHP}, \texttt{Java} and \texttt{JavaScript}. 
Each of the remaining 99 languages has a share of less than 1 percent and is ignored in the following.

\begin{figure}[htb]
  \centering
  \includegraphics[width=0.55\textwidth,page=2]{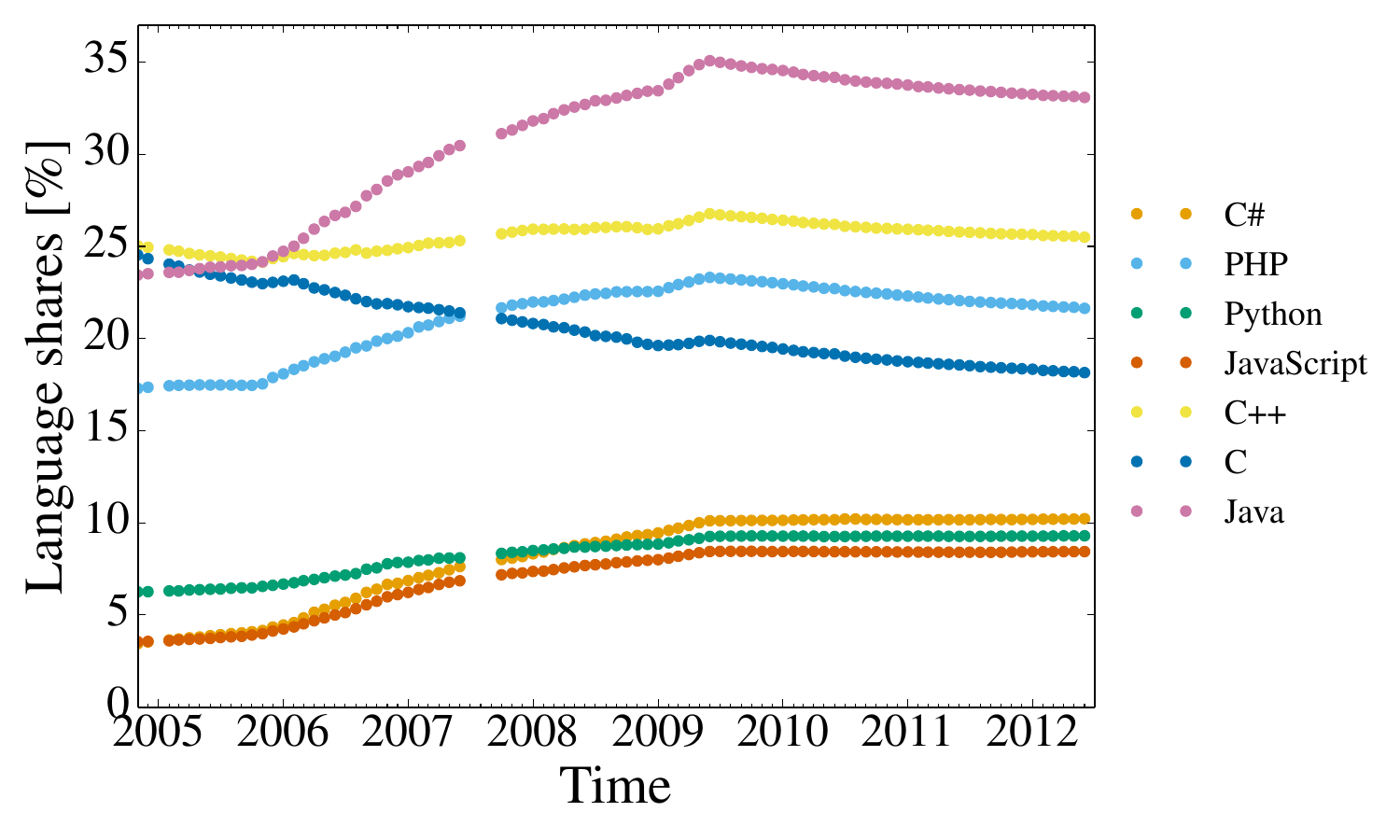}\hfill
  \includegraphics[width=0.44\textwidth,page=3]{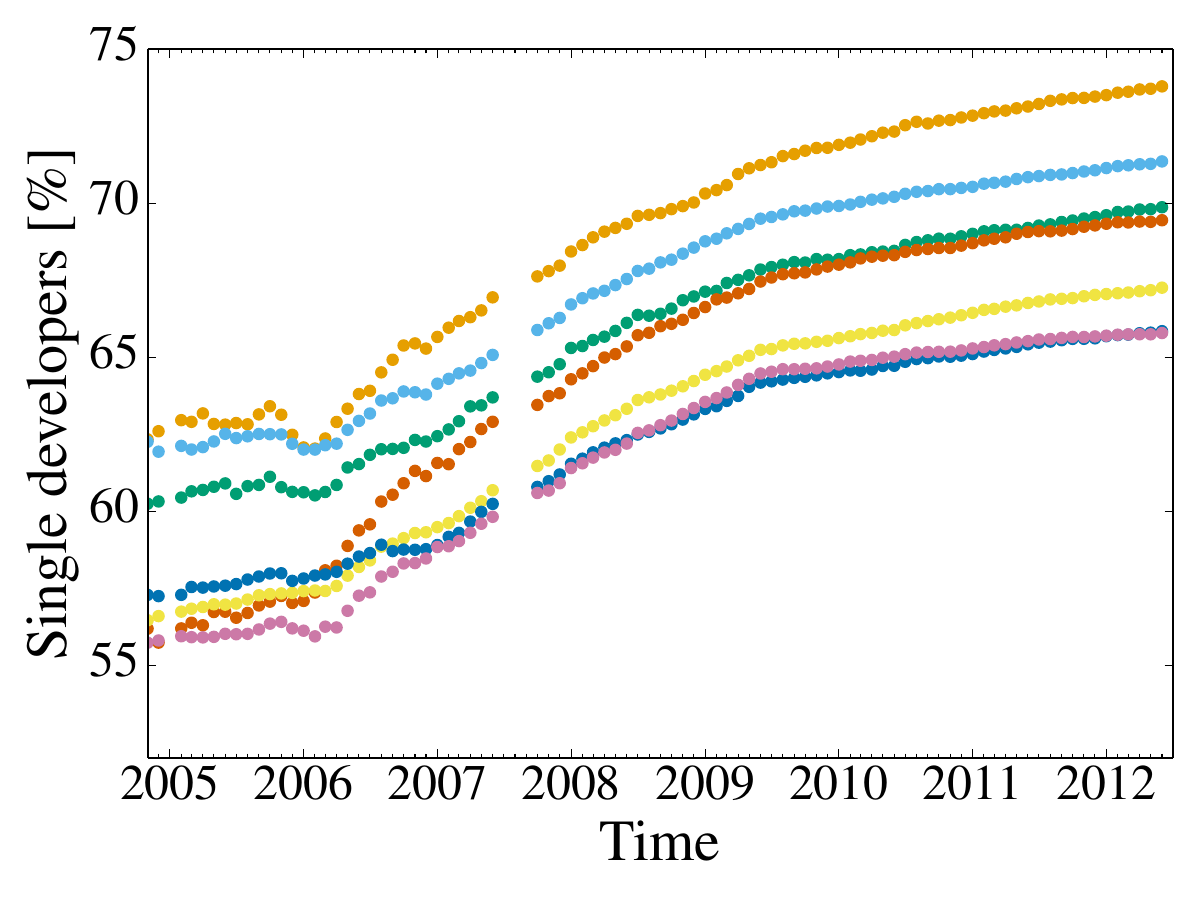}
  \caption{\textbf{(left)} Share of the 7 most popular programming languages (normalized to 100\%) over time measured in months, for all projects with available information about programming languages. 
\textbf{(right)} Share of single-developer projects (normalized for  each of the 7 most popular programming languages separately) over time measured in months.
}
  \label{fig:Programming Language 1}
\end{figure}

The importance of the major 7 languages changed considerably over time, as Fig.~\ref{fig:Programming Language 1} (left) shows.
Despite the fact that this refers only to a subset of projects, we can observe that 
\texttt{C} lost nearly 10\% market share  in 7 years (from 25\% down to 15\%), which is a loss of 40\% of its original total market share against the other 6 languages
even if the \emph{absolute} number of projects using \texttt{C} has increased. 
\texttt{Java}, on the other hand, increased its market share by about 10\%. But the largest shares are taken by 
\texttt{JavaScript} and \texttt{C\#}.

Figure~\ref{fig:Programming Language 1}(right) plots, for each of the 7 main programming languages, how the share of single-developer projects changes over time. 
We first note the trend towards more single-developer projects for \emph{all} of these 7 languages, but with noticeable language preferences.
In July 2012, 76\% of all projects using  \texttt{C\#} are single-user projects, followed by \texttt{PHP} with a share of 74\% single-developer projects and \texttt{Python} and \texttt{JavaScript} with 72\%.
I.e., developers who prefer to work on their own, have a clear preference for these languages.

\subsection{The bipartite network of developers and projects}
\label{sec:bipartite}

\begin{figure}[htb]
  \centering
  \includegraphics[width=0.9\textwidth]{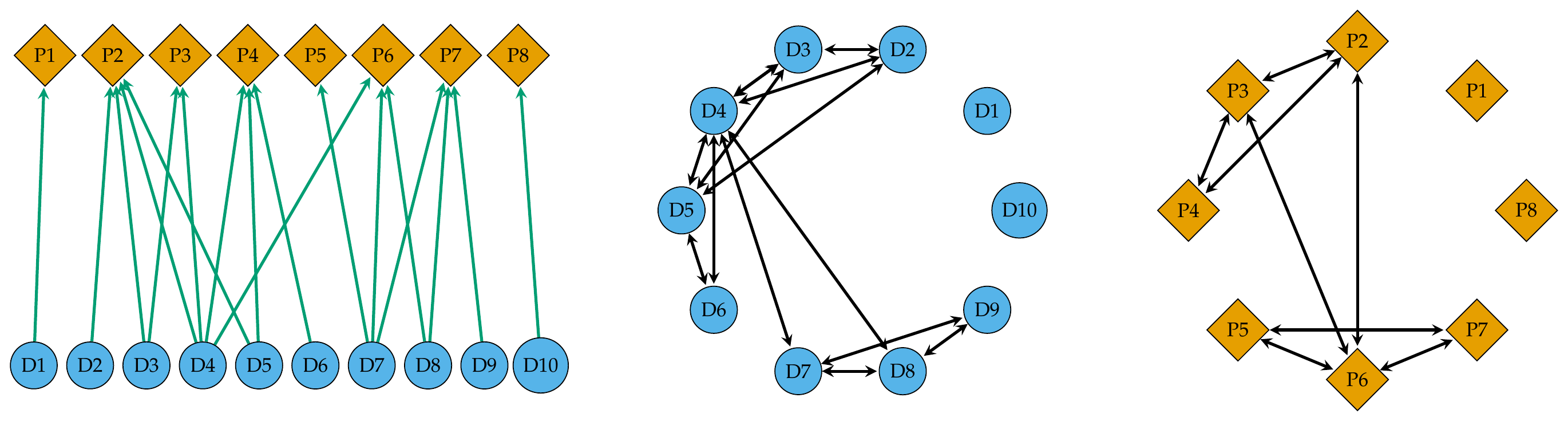}

  \caption{
\textbf{(right)} Example of a bipartite network where 10 developers labeled $D1,\cdots,D10$ contribute to eight projects labeled $P1,\cdots P8$. (\textbf{middle}) Projection of the network of developers (linked by common projects), \textbf{(right)} projection of the network of projects (linked by common developers).}
  \label{fig:Bipartite Network Example}
\end{figure}

We now take a closer look at the developers and their projects.
Both form a \emph{bipartite network}, i.e., a network where links exist between \emph{different} types of nodes.
As explained above, we consider a \emph{link} between a developer and a project if this developer has registered for the project regardless of her subsequent activity.
Thus, instead of a weighted network where the weight of the links reflects the contribution, in this paper we only consider an \emph{unweighted} network.
A sketch of this bipartite network is shown in Fig.~\ref{fig:Bipartite Network Example}, where 10 developers  contribute to eight different projects.
I.e., links between developers only exist through projects, and links between projects only through developers.

Nevertheless, we can draw two projections of this \emph{bipartite network} also shown in Fig.~\ref{fig:Bipartite Network Example}, one with respect to the \emph{developers} and one with respect to the \emph{projects}.
In these projection, a link between \emph{developers} appears if both of them contribute to the \emph{same} project, and a link between \emph{projects} appears if both of them have the same developer contributing.
We emphasize that the bipartite network and its projections are \emph{aggregated} over a given time interval, i.e., a link essentially reflects that two developers contributed to the same project in the same time interval (and not necessarily at the same time).

\begin{figure}[htb]
  \centering
  \includegraphics[width=0.49\textwidth,page=6]{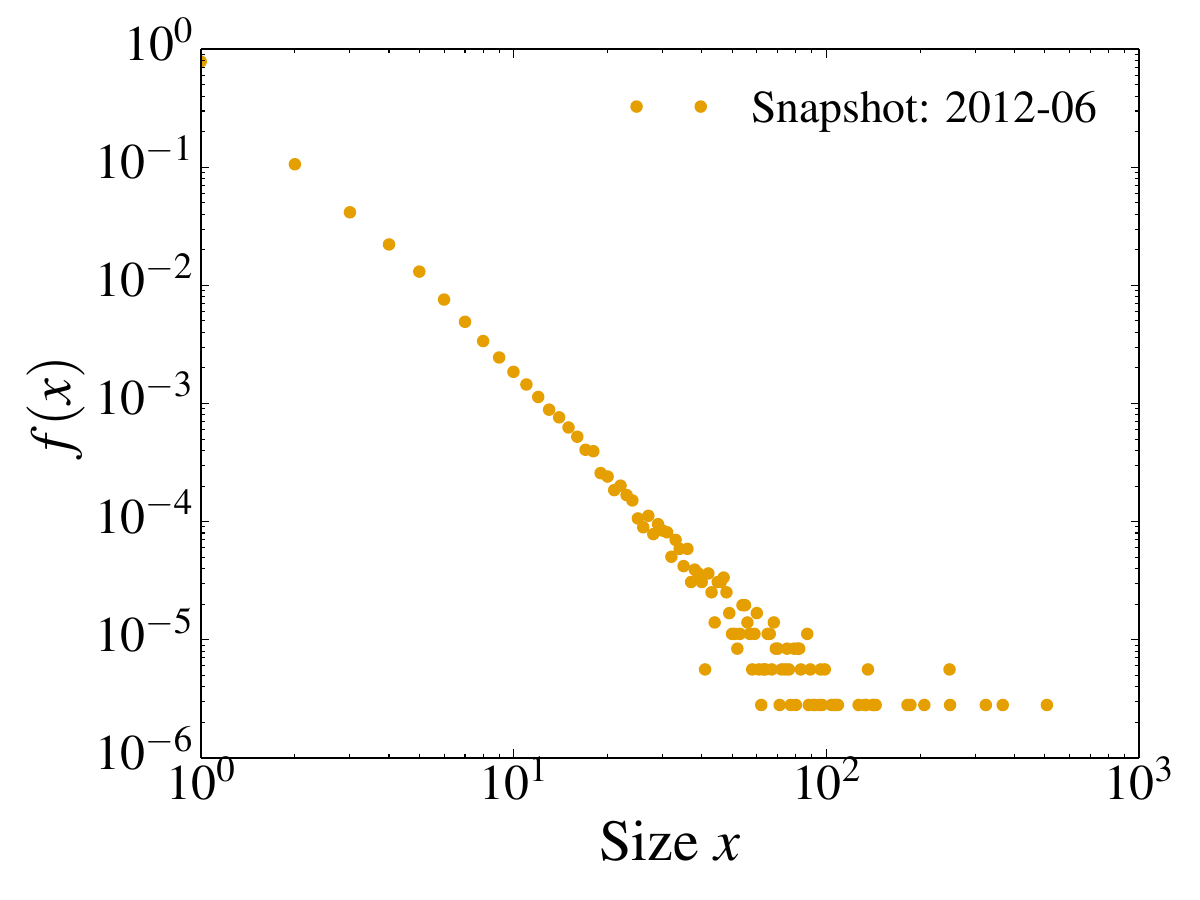}\hfill
  \includegraphics[width=0.49\textwidth,page=6]{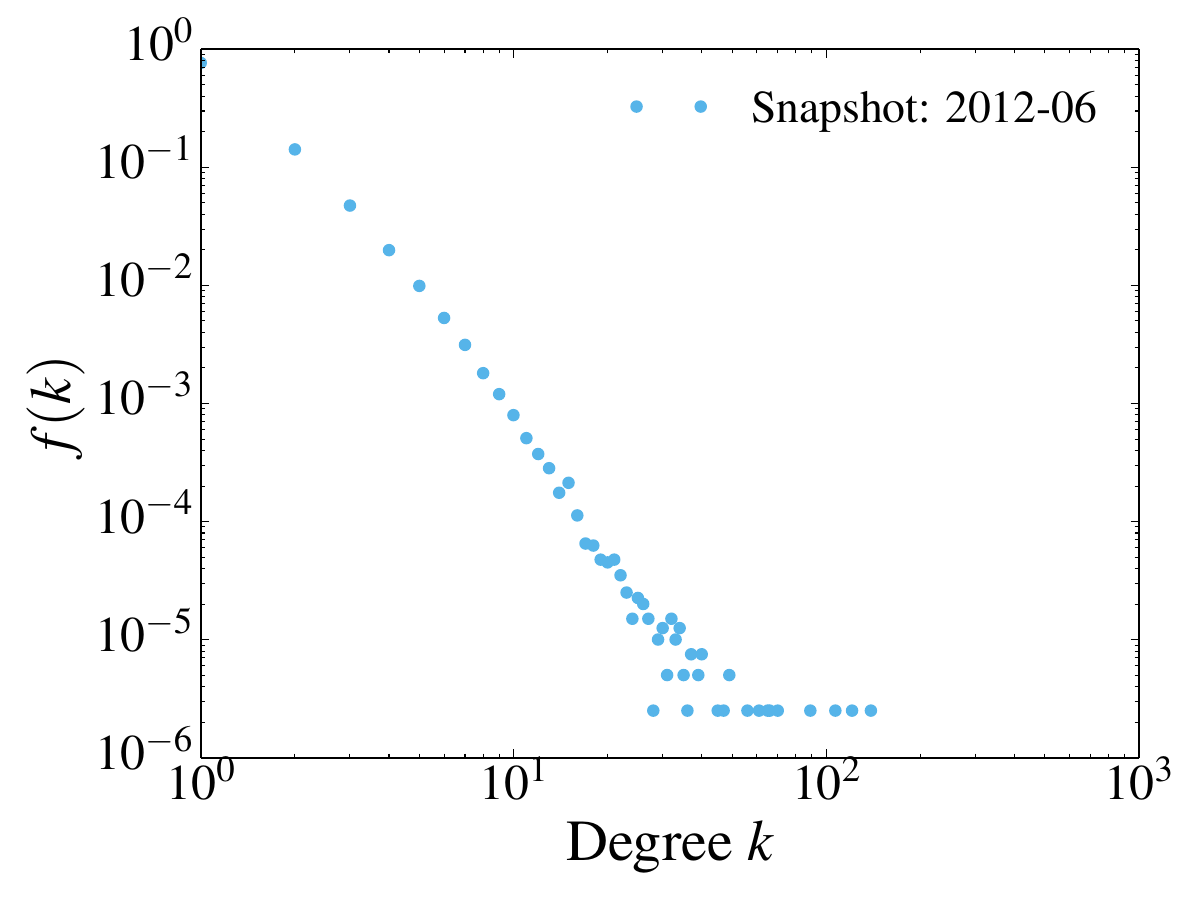}
  \caption{
\textbf{(left)} Degree (size) distribution of projects (i.e.,~number of developers per project), $f(x)$, \textbf{(right)} Degree distribution of developers (i.e.,~number of projects a developer contributes to), $f(k)$, for the monthly snapshot of  June 2012.
}
  \label{fig:CDF}
\end{figure}

Based on the aggregated description, we can define the \emph{degree} $k_{i}$ of a developer $i$ as the number of links she has, i.e., the total number of projects she was involved over that time period.
Likewise, we can also define the degree $x_{r}$ of a project $r$ as the total number of developers that contributed to this project over that time period.
$x$ is also called the \emph{size} of the project, as measured by the number of developers.
The network of developers can then be described by a \emph{degree distribution} $f(k)$ which gives the fraction of developers with degree $k$ in the population of all developers, during the observation period.
Likewise the degree distribution, later also called \emph{size distribution}, $f(x)$ gives the fraction of projects with $x$ developers, during  the observation period.
Both distributions are plotted in Fig.~\ref{fig:CDF} for the snapshot of 06/2012, which is the last snapshot of our dataset.

We observe that both are very skew distributions, reflecting the fact that there is a considerable probability to find projects of large sizes, or developers involved in very many projects. The distributions resemble known \emph{scale-free} distributions (such as power-law distributions), which indicates that there is no characteristic scale (size, number of projects) for projects or developers. In fact, these are \emph{not} pure power-law distributions (note the bend in the shape and a rather limited range), but  the specific type will be discussed in Sect. \ref{sec:disc}

\section{The growth of OSS projects: A microscopic perspective}
\label{sec:micro}

\subsection{Entry and exit dynamics}
\label{sec:projects}

\begin{figure}[htb]
  \centering
  \includegraphics[width=0.49\textwidth]{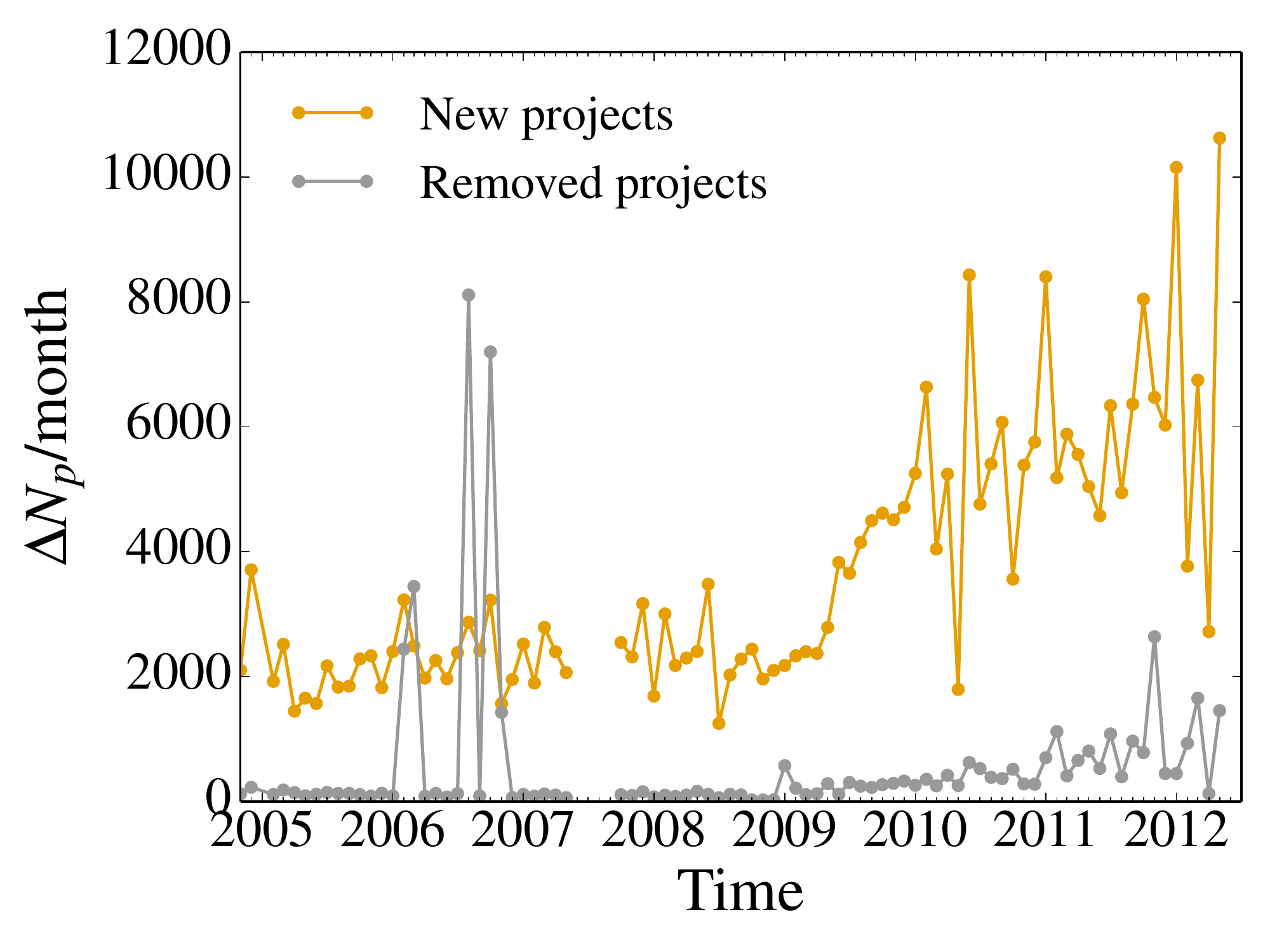}\hfill
  \includegraphics[width=0.49\textwidth]{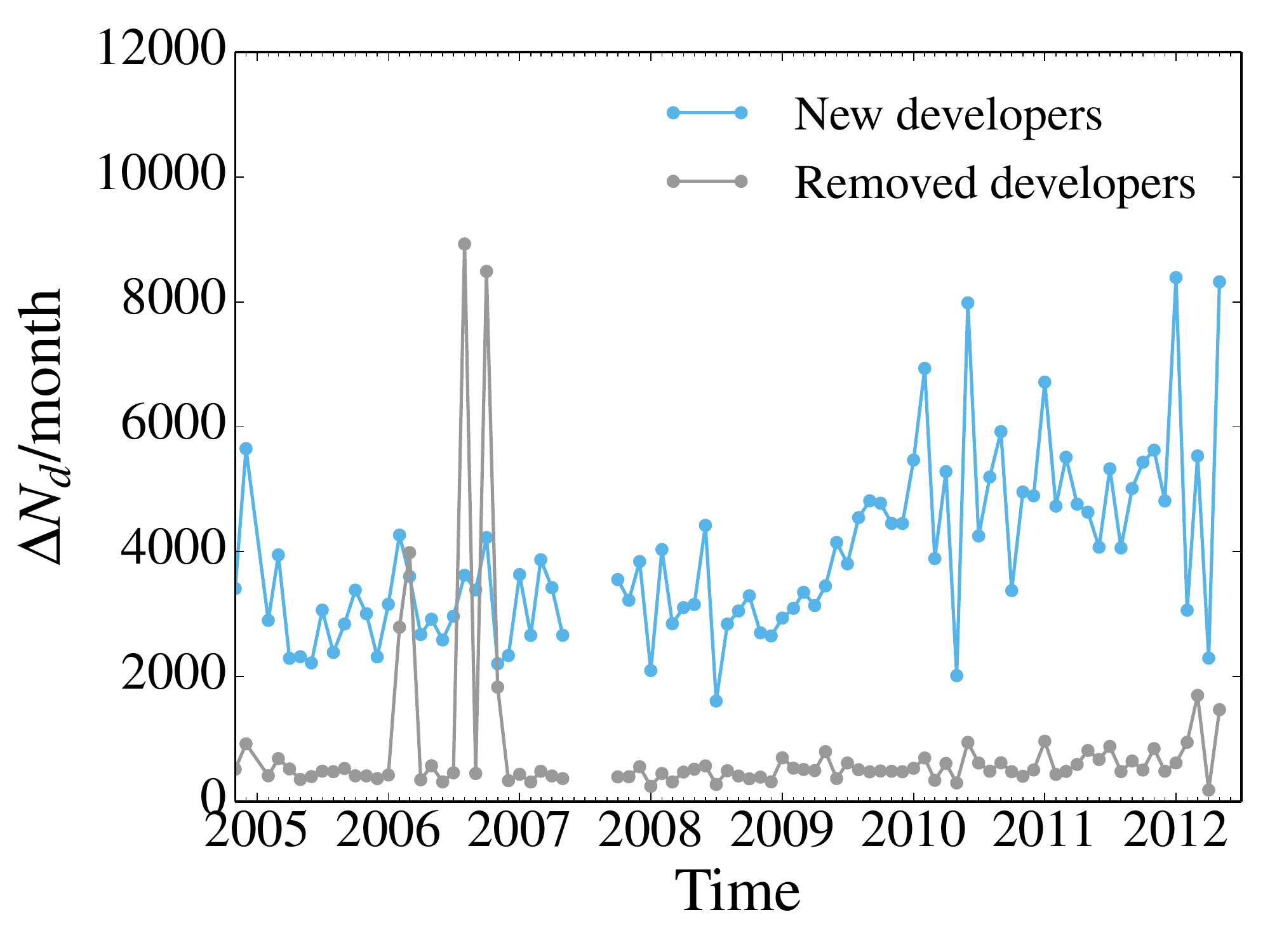}
  \caption{ 
(\textbf{left}) Number of new projects (yellow) and removed projects (grey) per month.
(\textbf{right}) Number of new developers (blue) and removed developers (grey) per month. 
}
  \label{fig:Project EnEX Interval=1}
\end{figure}

In this section, we analyze the dynamics of projects and of the developer community in more detail, by looking at the available data about birth and death of projects and entry and exit of developers, instead of the aggregated growth.
Figure~\ref{fig:Project EnEX Interval=1}(left) shows the number of \emph{new} projects per months, as well as the number of \emph{removed} projects per  month, while Fig.~\ref{fig:Project EnEX Interval=1}(right) shows the corresponding numbers for \emph{developers}.
We call the underlying processes ``entry'' and ``exit'' of projects or developers, respectively. 

We can immediately observe that the number of entry events largely exceeds the number of exit events, for each month, both for projects and developers, with the exception of the large exit spikes observed in 2006-2007 because of the project clean-up initiated by SF (see Sect. \ref{sec:data}).
The reason for the dominance of the entry dynamics in normal time periods is that most projects or developers are not really \emph{removed} if they become \emph{inactive}. 
In fact, it is not trivial to determine whether a project or a developer is really inactive. 
Often, the activity just decreases considerably, but does not cease to exist.
Also, the fact that there is no activity in a given time period does not imply that there will be also no activity in the future. 
We have discussed this question in detail in \citep{github}.
In this paper, we do not speculate about inactivity and just take the computed exit rates as a matter of fact.
For the modeling in Section \ref{sec:macro}, we take advantage of their very low numbers and will simply neglect the exit dynamics.

Eventually, we also note the occasional large fluctuations in the exit rates, both for projects and developers. 
These are the results of extraordinary efforts by SF to clean up the project and developer base, e.g., by testing and implementating the new Autopurge System after turning off the old one (see Section \ref{sec:data}). 
During and shortly after this switch, either extremely higher or lower numbers of projects and developers were detected as  inactive and removed.

\begin{figure}[htb]
  \centering
  \includegraphics[page=1,width=0.49\textwidth]{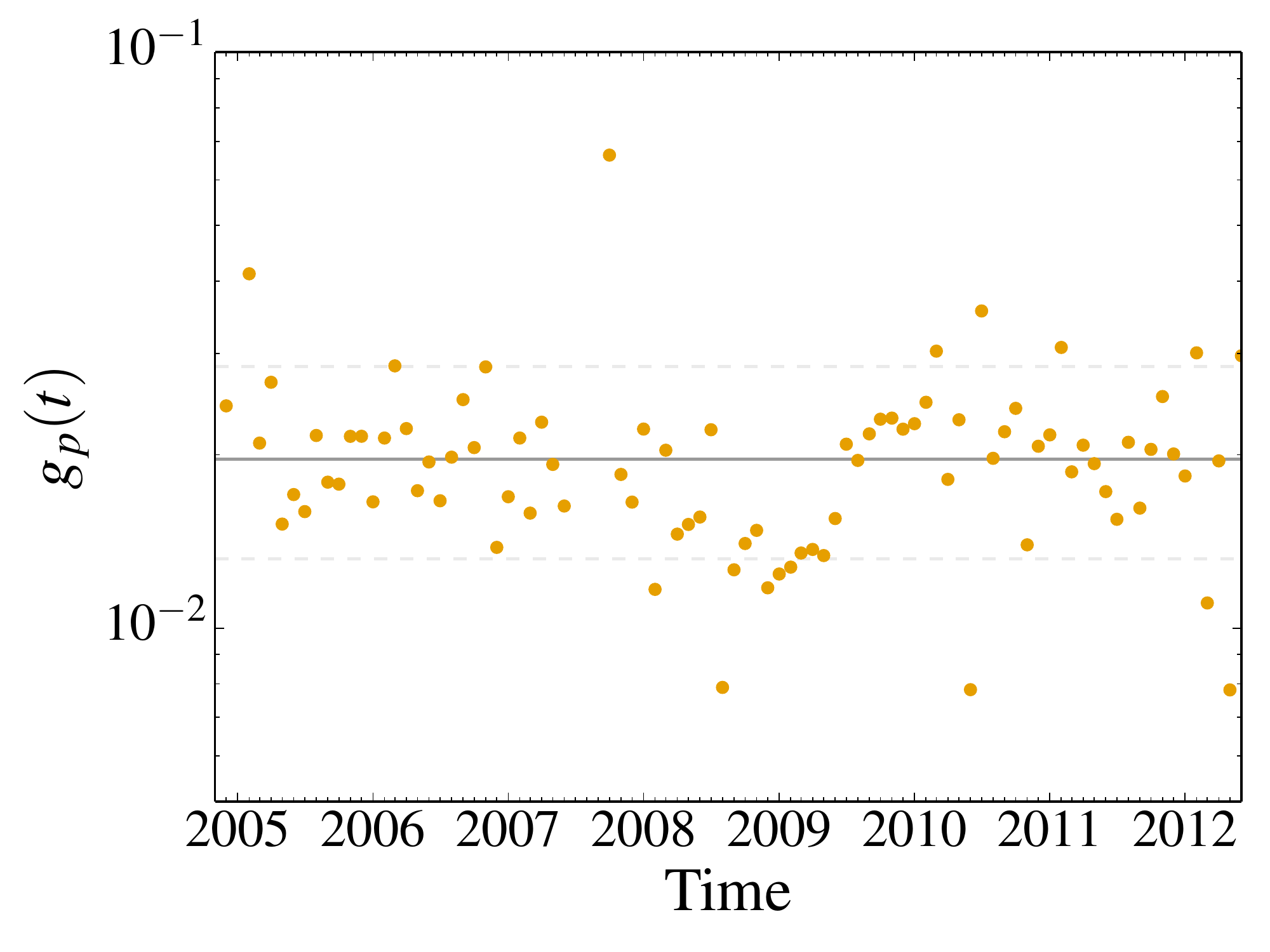}\hfill
  \includegraphics[page=2,width=0.49\textwidth]{P-prob_new_new.pdf}
\caption{Relative monthly entry rate of projects, $g_{p}(t)$ \textbf{(left)}, and of developers, $g_{d}(t)$  \textbf{(right)}.
The horizontal lines represent the median of all the values in the time series (solid line) and the dashed lines, $10\%$ and $90\%$ quantiles.
The values read for $g_{p}$: 10\%:  0.0131 , median:  0.0196, 90\%:  0.0284, and for $g_d$: 10\%:  0.0118, median:  0.0164, 90\%:  0.0232
}
  \label{fig:relative}
\end{figure}

The second important observation is the \emph{growth} of the monthly \emph{entry rate}s over time, both for projects and developers (indeed, for projects, we could also note an increase of the exit rates over time). 
High occasional fluctuations might result from seasonal factors (holidays) or high media attention.
This growth on average is in line with the exponential growth observed both for projects and developers on the aggregated level as discussed in Section \ref{sec:growth}. 
The \emph{law of proportionate growth} tells us that SF, for the observed time interval, became more attractive the bigger it was.  
Hence, the monthly entry  rates shall depend on the current numbers of projects or developers, respectively. 
Figure~\ref{fig:relative} plots these \emph{relative monthly entry rates}
\begin{equation}
  \label{eq:relative}
  g_{p}(t)=\frac{N_{p}(t)-N_{p}(t-\Delta t)}{N_{p}(t)}\;;\quad
  g_{d}(t)=\frac{N_{d}(t)-N_{d}(t-\Delta t)}{N_{d}(t)} 
\end{equation}
both for projects and developers. 
We note that, despite some considerable fluctuations, they tend to vary around long-term stationary values $\bar{g}_{p}$, $\bar{g}_{d}$ in a first order approximation (i.e., we do not discuss a non-linear dependency on $N$). 

\subsection{Size dependent growth rates of projects}
\label{sec:developers}

So far, we have discussed the \emph{law of proportionate growth} on the aggregated level, both with respect to the absolute numbers $N_{d}$, $N_{p}$, and $K$, Eq. (\ref{eq:exp}), and the relative growth rates, $g_{d}$, $g_{p}$, Eq. (\ref{eq:relative}). 
But we can also refer to the individual project level, to verify this dynamics. 

Recall that the size $x_{r}$ of a project $r$ is defined by the number of developers contributing to it ($x_{r}$ was also called the degree of the project because, in the bipartite network, links exist between developers and the project). 
Then, the growth dynamics on the individual project level reads as: 
\begin{equation}
\frac{x_r(t+\Delta t) - x_r(t)}{\Delta t}\propto x_{r}^{\gamma}(t)
\label{eq:xrt}
\end{equation}
If $\gamma = 1$, we have a growth strictly proportional to size, which is also known as preferential attachment in network theory, i.e., nodes (projects) receive new links (developers) proportional to the number of existing links. 
$\gamma>1$ would indicate a super-linear growth. 

\begin{figure}
  \centering
  \includegraphics[page=1,width=0.49\textwidth]{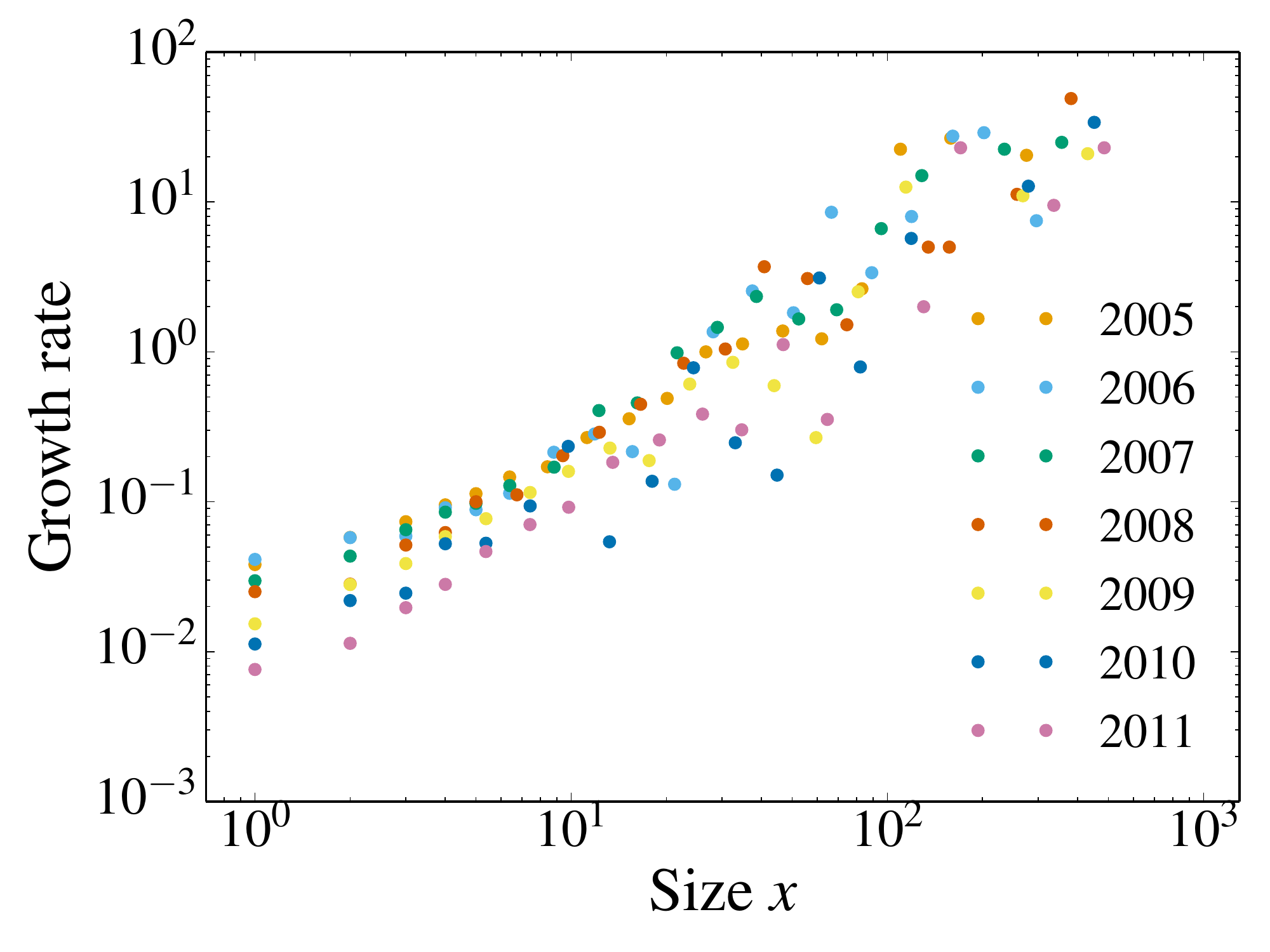}\hfill
  \includegraphics[page=3,width=0.49\textwidth]{P-growth_new.pdf}
  \caption{\label{fig:Growth}
\textbf{(left)} Averaged annual growth rate $\bar{g}(x)$ over project size $x$ (measured by the number of developers) for each year separately. \textbf{(right)}  Exponent $\gamma$, Eq. (\ref{eq:xrt}) for each year, where  
error bars indicate the standard error.
}
\end{figure}

As we have seen on the aggregated level, growth rates heavily fluctuate for each month. 
Therefore, for the individual project level, we choose the time window $\Delta t$=12 months, i.e., large enough to cancel out these fluctuations. 
We then compute the average growth rate $\bar{g}(x)$ of projects with similar size $x$ for each year, which is shown in Fig.~\ref{fig:Growth} (left). 
We verify that the annual growth rate indeed increases with the size of the project, as described in Eq. (\ref{eq:xrt}), and we barely notice differences in this dependence for different years. 

For a closer inspection of the law of proportionate growth, we estimate the exponent $\gamma$, Eq. (\ref{eq:xrt}), from the data separately for each year. 
We find that $\gamma$ varies indeed between 1.23 and 1.35 during the seven years period, hence the growth is slightly super-linear for all times. 
However, because $\gamma$ is very close to 1, we can still argue that the law of proportionate growth approximately holds, but there is a higher order dependence on the project size $(\sim x^{1/4})$ which may enter the proportionality factor $\beta$, i.e., $\dot{x}=\beta(x)\, x$.

\section{A dynamic model of project growth}
\label{sec:macro}

\subsection{Dynamic assumptions}
\label{sec:rate}

In the following, we focus on the dynamics at the \emph{project} level, only.
The number of projects of a given size $x$ (measured by the number of developers contributing to the project) at a given time $t$ is given by $n(x,t)$.
Each project of size $x$ at time $t$ belongs to the same \emph{size class} $Y_{t}=x$.
Time $t$ is assumed to be discrete (measured in months).
The total number of projects, $N_{p}(t)$, and the total number of developers, $N_{d}(t)$, contributing to projects are defined as:
\begin{equation}
  \label{eq:total}
  N_{p}(t)=\sum_{x=1}^{x_{\mathrm{max}}}n(x,t)\;;\quad
  N_{d}(t)=\sum_{x=1}^{x_{\mathrm{max}}}x\ n(x,t)
\end{equation}

For our dynamic assumptions, we follow the model of Simon \citep{simon1955} for the entry of new firms and the growth of existing firms. 
That means in our model the \emph{entry of new developers} is assumed to be the only source for (i) establishing \emph{new projects} and (ii) \emph{enlarging} existing ones.
Precisely, we neglect the possibility that also established developers already involved in other projects found a new project. 
This is supported by the empirical finding that most developers are only involved in one project, as the degree distribution of developers in Fig. \ref{fig:CDF}(right) shows.
But we will come back on the validity of this assumption in Sect. \ref{sec:disc}. 

We further neglect \emph{fragmentation}, or \emph{fork} processes, i.e., an existing larger project splits up into two (or more) smaller projects, which could be also seen as the foundation of a new project of size $x\geq 1$.
 We argue that such events exist but are comparably rare so that we cannot suffiently calibrate our model against such data, and simply neglect these events.

The empricial finding of Fig. \ref{fig:OSS Growth Vs. Time} tells us that the number of developers has increased exponentially.
We can include this in three different ways: (a) choosing a \emph{linearly increasing} number of new entrants per time interval, (b) rescaling the time interval linearly such that the number of new entrants per time interval is constant, (c) replacing time $t$ simply by the total number of developers $N_{d}$.
We have chosen the latter as the most elegant way.
Hence from $N_{d}(t)\propto \exp{(\omega t)}$, we find the transformation $t\to \ln{N_{d}}/\omega$.
From now on $N_{d}\equiv N$ measures time in discrete steps, $N$, $N+1$.... 

For the change of $n(x,t)$, we discuss the following processes:
\begin{enumerate}
\item \emph{A new project is founded:} Here the assumption is that the project starts in the smallest size class $x=1$.
There is   a certain (conditional) probability
  \begin{equation}
    P_{0,1}^{N+1}(Y_{N+1}=1|Y_{N}=0)=p_{0}   \quad (x=1)
    \label{eq:p01}
  \end{equation}
that we find in the next time step $N+1$ a new project of size 1 (where its former size 0 indicates that the project did not exist yet at time $N$).
This probability is denoted as $p_{0} \in(0,1)$ and assumed to be constant in time, except for the very first time step $N=0$ at which no projects exist yet.
So one has to be founded with certainty:
\begin{equation}
P_{0,1}^{1}=(Y_{1}=1|Y_{0}=0)=1
\label{eq:start}
\end{equation}
i.e., we start the dynamic process with one project that is of the smallest possible size 1.
Because at each time step only one new developer enters, the largest possible size of any project cannot be larger than $N$, i.e.,  we set $x_{\mathrm{max}}=N$.
\item \emph{An established project grows:} Here the assumption is that the project only grows by attracting \emph{one new developer}at a time.
This event is described by the probability
  \begin{equation}
    \label{eq:pn1}
          P_{x-1,x}^{N+1}(Y_{N+1}=x|Y_{N}=x-1)=K(N)\ (x-1)^{\alpha}\ n(x-1,N) \quad (x=2,...,N)
  \end{equation}
I.e., the (conditional) probability of a project in size class $(x-1)$ to grow at time $N$ is proportional to the number of projects in that size class, $n(x-1,N)$.
However, the new developer may have a \emph{preference} for larger or smaller projects, i.e., the probability to choose from size class $(x-1)$ is also proportional to $(x-1)^{\alpha}$.
$\alpha=0$ would recover the case of \emph{no size preference}, which was discussed, e.g., in \citep{sutton}.
$\alpha=1$ would be a preference directly proportional to the existing size, which was discussed in \citep{simon1955} to cope with Gibrat's \emph{law of proportionate growth}.
In the following, because of analytical tractability we will only consider the case $\alpha=1$.
Its empirical evidence is discussed in the next section, while in Sect. \ref{sec:disc} consequences for different values of $\alpha$ are discussed. 

$K(N)$ is a proportionality constant that has to satisfy the condition that all probabilities sum up to 1.
Using $\alpha=1$ from now on, we find:
\begin{equation}
  \sum \limits_{x=1}^{N}K(N)\ x\ n(x,N)+p_{0}\stackrel{!}{=}1
  \label{eq:probsum}
\end{equation}
Because of Eq. (\ref{eq:total}), $\sum \limits_{x=1}^{N}x\, n(x,N)=N$, and we get from Eq. (\ref{eq:probsum})
\begin{equation}
K(N)\ N +p_{0}= 1 \quad \Rightarrow \quad K(N)=\frac{1-p_0}{N}
  \label{eq:factork}
\end{equation}
Note that this would not hold if $\alpha \neq 1$.
Equation \ref{eq:factork} allows us to rewrite Eq. (\ref{eq:pn1}) as
\begin{equation}
  P_{x-1,x}^{N+1}(Y_{N+1}=x|Y_{N}=x-1)= \left(1-p_{0}\right)\frac{(x-1)\ n(x-1,N)}{N},\quad x=2,...,N
  \label{eq:a1}
\end{equation}
\end{enumerate}
Our kinetic assumptions as seen from the perspective of the developer, are summarized as follows: at each time step $N+1$, \emph{one new developer arrives}.
This developer has \emph{two options}: (i) with probability $p_{0}$ she chooses to found a \emph{new} project, (ii) with probability $(1-p_{0})$, she chooses to join one of the projects that exist at time $N$, i.e., $N_{p}(N)$.
 Without any preference for larger projects, she will choose a project from size class $x$ with a probability $(1-p_{0})n(x,N)/N$.
But with the assumed size preference, $\alpha=1$, the proportional weight $x$ comes into account. 

We emphasize that some of the dynamic processes one could think of are deliberatively neglected, e.g., we neglect that several developers join one or different projects during the same time interval (which can be solved by changing the time resolution).
 More importantly, we also neglect that developers switch between projects, i.e., some projects lose and some projects gain in developers, but the total number of developers does not change.
Such dynamics can be seen as reallocation processes among projects, and are neglected here.

Further we do not consider that \emph{existing projects shrink}, i.e., loose in size if developers leave.
Since we have opted out reallocation processes, it would mean that developers become inactive.
Again, there is empirical evidence for this process (see Fig. \ref{fig:Project EnEX Interval=1}).
But the number of developers exiting is (a) rather constant in time, and (b) much smaller than the number of new developers arriving.
Therefore, we will consider this in our model as a rescaling of the arrival rate of new developers and will not explicitly model the shrinking process.

Eventually, we also do not consider that an \emph{established project ceases to exist}, either.
Such processes can happen in two ways: (a) the project goes extinct, and (b) two existing projects \emph{merge} into a new one, with a larger project size.
Again, in our model we neglect both processes.
Projects often become \emph{inactive}, but are rarely deleted, and for \emph{mergers and acquisitions} the same argument as for the project forks apply.

With these considerations the total number of projects at time $N$ is given by:
\begin{equation}
  \sum\limits^N_{x=1} n(x,N) \approx1+p_0(N-1) = Np_0
  \label{eq:number}
\end{equation}
The first term, 1, results from the fact that there exists a new project in the first time step.
Then, during every time step from $N=2$ up to $N$ a new project appears with probability $p_0$.
Hence, if $p_{0}\ll 1$ and $N$ is large, this gives approximately 
$1+p_0(N-1)$, which is $Np_0$.

\subsection{The rate equation}
\label{sec:rate}

We now start formalizing the above assumptions by developing a rate equation for the relevant quantity $n(x,N)$.
This can change by two processes: (a) \emph{gain:} a project of size $x-1$ is chosen by the developer and thus advances to the next size class, leading to an \emph{increase} in $n(x,N)$ (b) \emph{loss:} a project of size $x$ is chosen by the developer and thus advances to the next size class, leading to a \emph{decrease} in $n(x,N)$.
\begin{equation}
 n(x,N+1) -n(x,N)=(1-p_{0})\left[\frac{(x-1)\, n(x-1,N)}{N}-\frac{x\, n(x,N)}{N}\right] \quad (x=2,...,N+1)
  \label{eq:difference1}
\end{equation}
The LHS of Eq. \eqref{eq:difference1} describes the  \textit{net inflow} of projects into size class $x$. 
Similarly we get for the number of new projects at time $N+1$:
\begin{equation}
 n(1,N+1) -n(1,N)=p_{0}-(1-p_{0})\frac{n(1,N)}{N} \label{eq:difference2}
\end{equation}
The \emph{gain} comes from funding new projects with a probability $p_{0}$, whereas the \emph{loss} results from the fact that a project of size 1 grows into size 2. 

In the following we will only consider ``steady-state'' distributions, i.e., we assume that each size class grows \textit{proportionally} with $N$:
\begin{equation}
  \frac{n(x,N+1)}{n(x,N)}=\frac{N+1}{N} \quad \forall \quad x,N \quad (\mathrm{if}\;\; x<N) 
  \label{eq:steady}
\end{equation}
With $\Delta n(x,N):=n(x,N+1)-n(x,N)$, we rewrite Eqs. \eqref{eq:difference1}, \eqref{eq:difference2} as: 
\begin{align}
  \Delta n(x,N)&=(1-p_{0})\left[\frac{(x-1)\, n(x-1,N)}{N} -\frac{x\, n(x,N)}{N}\right] \quad (x=2,...,N+1) \nonumber \\
\Delta n(1,N)&= p_{0}-(1-p_{0})\frac{n(1,N)}{N}
  \label{eq:stst2}
\end{align}
From Eq. \eqref{eq:steady} it follows that:
\begin{align}
   n(x,N+1)  =(1+\frac{1}{N})n(x,N) \;; \quad 
 \Delta n(x,N) =\frac{n(x,N)}{N}
  \label{eq:folgen}
\end{align}
Plugging Eq. \eqref{eq:folgen} in Eqs. \eqref{eq:difference1}, \eqref{eq:difference2}, we have:
\begin{align}
    & \Delta n(x,N)=\frac{n(x,N)}{N}=(1-p_{0})\left[\frac{(x-1)\, n(x-1,N)}{N}-\frac{x\, n(x,N)}{N}\right], \;\; x=2,...,N \nonumber \\
    & \Delta n(1,N)=\frac{n(1,N)}{N}=-p_{0}- (1-p_{0})\frac{n(1,N)}{N}
  \label{eq:solve1}
\end{align}
which simplifies to the following set of equations:
\begin{align}
 0 & = (1-p_{0})(x-1)\,n(x-1,N)-(1-p_{0})x\,n(x,N)-n(x,N) \nonumber \\
 0 & = Np_{0} -(1-p_{0})n(1,N)-n(1,N)
  \label{eq:solve2}
\end{align}

\subsection{The size distribution of projects}
\label{sec:distr}
In order to  solve Eq. \eqref{eq:solve2}, we define the new parameter $\rho$
\begin{equation}
\label{rho}
  \rho=\frac{1}{1-p_{0}}
\end{equation}
where $1<\rho<\infty$, because of $p_{0} \in (0,1)$.
To interpret $\rho$ \citep{simon1955}, we keep in mind that $p_{0}$ actually decides how much of the growth (one developer per time unit) is spent on new projects as compared to established projects.
In Sect. \ref{compare} we will test this relation against our empirical data. 

From Eq. (\ref{eq:solve2}), we find for the stationary solution for $n(1,N)$ (denoted by $^{*}$)
\begin{equation}
  n^*(1,N)=\frac{Np_{0}}{2-p_{0}}=\frac{\rho}{\rho+1}Np_{0}
  \label{eq:star1}
\end{equation}
whereas we find for the stationary solution of $n(x,N)$:
\begin{equation}
  n^{*}(x,N)=\frac{(1-p_{0})(x-1)}{1+(1-p_{0})x}n^{*}(x-1,N)=\frac{(x-1)}{\rho+x}n^{*}(x-1,N)
  \label{eq:solve1.1}
\end{equation}
We can solve this equation in an iterative manner, to find:
\begin{equation}
  n^*(x,N)=\frac{(x-1)}{(\rho+x)}\ \frac{(x-2)}{\left(\rho+(x-1)\right)}\ ...\ \frac{1}{(\rho+2)}\ n^*(1,N) \quad (x=2,...N)
  \label{eq:iteration2}
\end{equation}
To further compact this expression, we make use of the  so-called \emph{Gamma function} $\Gamma(z)$ with the property $\Gamma(z+1)=z\Gamma(z)$ that, for integer $z$, results in: 
\begin{equation}
  \Gamma(z)=(z-1)!
  \label{eq:gamma1}
\end{equation}
The denominator of Eq. (\ref{eq:iteration2}) can then be expressed as:
\begin{equation}
  \Gamma(x+\rho+1)=(x+\rho)\Gamma(x+\rho)=(x+\rho)(x+\rho-1) ... (2+\rho)\Gamma(\rho+2)
  \label{eq:gamma2}
\end{equation}
With this and the expression for $n^*(1,N)$, Eq. \eqref{eq:star1}, we can rewrite Eq. \eqref{eq:iteration2}:
\begin{align}
n^*(x,N)&=\frac{\Gamma(x)\Gamma(\rho+2)}{\Gamma(x+\rho+1)}f^*(1,N) \nonumber \\ &=(\rho+1)\frac{\Gamma(x)\Gamma(\rho+1)}{\Gamma(x+\rho+1)}f^*(1,N)=\rho \mathcal{B}(x,\rho+1)Np_{0}
  \label{eq:betafunc}
\end{align}
where ${\Gamma(x)\Gamma(\rho+1)}/{\Gamma(x+\rho+1)}=\mathcal{B}(x,\rho+1)$ is the \textit{Beta function}.

The corrected, normalized \emph{Yule-Simon distribution}, which holds also for $x=1$, is then:
\begin{equation}
  f(x,N):=\frac{n^*(x,N)}{Np_{0}}=\rho \mathcal{B}(x,\rho+1) = \frac{\rho\,\Gamma(\rho+1)}{(x+\rho)^{{\rho+1}}}  \quad (x=1,2,..)
  \label{eq:pdistr}
\end{equation}
For large $x$ we get for $f(x,N)$:
\begin{align}
  f(x,N)=\rho\mathcal{B}(x,\rho+1)\approx \rho x^{-(\rho+1)}, \quad x\rightarrow \infty
\label{eq:approx}
\end{align}
which has the form of a power law if $x$ is large enough.
Therefore, the power law approximates the Yule-Simon distribution only in its upper tail.
To get Zipf's law, one has to assume $p_{0} \rightarrow 0$, as $\rho={1}/{(1-p_{0})}\approx 1$.
However, as noticed, e.g. by \citep{Krugman1996self}, for $p_{0}\rightarrow 0$  the convergence to the steady-state is infinitely slow.

\section{Discussion}
\label{sec:disc}

\subsection{Comparison with the Yule-Simon distribution}
\label{compare}

We have now a theoretical prediction for the size distribution of projects, Eq. (\ref{eq:pdistr}), and we have the respective empirical data for different years. 
Therefore, as a first step, we evaluate the kind of distribution that was already plotted in Fig. \ref{fig:CDF}. 

Before doing so, we have to argue whether the theoretical prediction and the empirical data really describe the same kind of projects. 
Our theoretical model is based on the assumption that \emph{all} projects entering the system are \emph{potentially} available to grow in the number of developers, i.e., developers can simply join them. 
This, however, cannot be confirmed for all single-developer projects listed in the database. 
Here, we have to consider that developers host their projects on SF not just to invite collaboration, but for various reasons, e.g. for archival purposes or just for distribution. 
While we cannot access the intrinsic reasons for a project to be hosted on SF, we argue that all new projects appearing on SF every month can be divided into \emph{two classes}: (a) \emph{collaborative} projects, i.e., projects that are meant to grow also by the contribution of other developers joining the project, and (b) \emph{non-collaborative} projects, that are not aimed at attracting other developers and thus do \emph{not} grow in size as measured by the number of developers, but maybe grow in their lines of code submitted by the project holder.
I.e., we conjecture that there is a sizable number of single-developer projects that, from their very beginning, are not captured by our model that applies only to \emph{collaborative} projects, i.e., projects with the potential to grow in the number of developers.

Consequently, when comparing the predicted size distribution with empirical data, we have to take into account that $f(1,N)$, i.e., the normalized density of projects of size 1, will need to be \emph{corrected}, to subtract the \emph{non-collaborative} projects, $f^{\cancel{c}}(1,N)$ (where $\cancel{c}$ stands for non-collaborative)  and to consider only the \emph{collaborative} ones, $f^{c}(1,N)$ . The procedure for this necessary correction will be described further below. The resulting corrected size distribution will then be indicated by ${f}^{c}(x,N)$. 
 
As the null hypothesis for the size distribution $f(x,N)$, we test for the Yule-Simon distribution for which 
the maximum likelihood of the parameter $\rho$ can be computed numerically \citep{Garcia20118560}.
We perform a Kolmogorov-Smirnov (KS) test to determine the significance level ($p$-value) for which the empirical distribution matches the Yule-Simon distribution. 
$p=0$ means that the two distributions do not match under any circumstances.  
The higher the $p$-value, the more likely it is that the null hypothesis cannot be \emph{rejected}.
That means we cannot exclude that the Yule-Simon distribution is the right distribution, but there might be also other candidate distributions that could be considered (which we abstain from). 

Applying the KS test in its simplest form to skew distributions usually results in very high $p$-values, simply because the mass of the distribution is mostly concentrated in the head while the tail is weighted less. 
The problems resulting from this naive approach have been discussed in detail in \citep{Clauset2009}. 
These authors also propose a more reliable, but computationally more demanding, goodness-of-fit test suitable for heavy-tailed distributions which we adopt here. 

We first turn to the degree distribution of \emph{developers}, an example of which is shown in Fig. \ref{fig:CDF}(right). 
Our goodness-of-fit test reveals that the, rather steep, ``broad'' developer degree distribution,  $f(k)$, does \emph{not} follow a Yule-Simon distribution ($p$=0).
But since we never made a hypothesis about this and did not develop a model for it, we just take this as a fact.

\begin{figure}
  \centering
\includegraphics[page=1,width=0.49\textwidth]{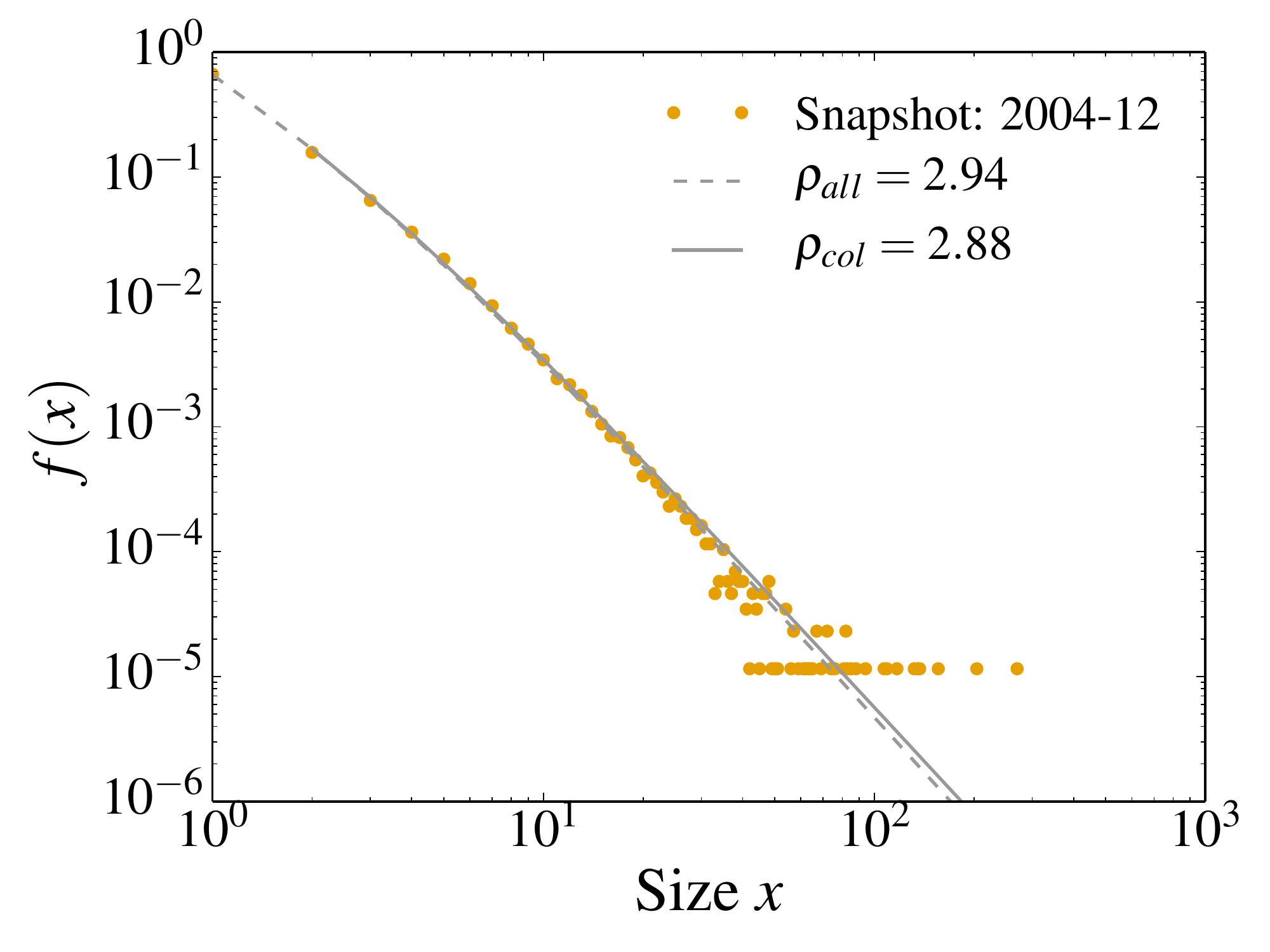}\hfill
  \includegraphics[page=2,width=0.49\textwidth]{P-yule-fit-SINGLES_OUT.pdf}\\
  \includegraphics[page=3,width=0.49\textwidth]{P-yule-fit-SINGLES_OUT.pdf}\hfill
  \includegraphics[page=4,width=0.49\textwidth]{P-yule-fit-SINGLES_OUT.pdf}
  \caption{\label{fig:YuleFit} 
Project size distribution $f(x)$ for different monthly snapshots. 
For each snapshot, a fit (dashed gray line) of the Yule-Simon distribution is plotted, for which the parameter $\rho_{\mathrm{all}}$ was numerically obtained. The goodness-of-fit test however rejects the hypothesis that the Yule-Simon distribution fits the empirical one for most of the snapshots.  
A second fit (solid gray line) of Yule-Simon distribution, for which the value for single-developer projects is taken as unknown, latent variable is also plotted, for which the same hypothesis cannot be rejected for the most of the snapshots.
}
\end{figure}

\begin{figure}
  \centering
  \includegraphics[width=0.49\textwidth]{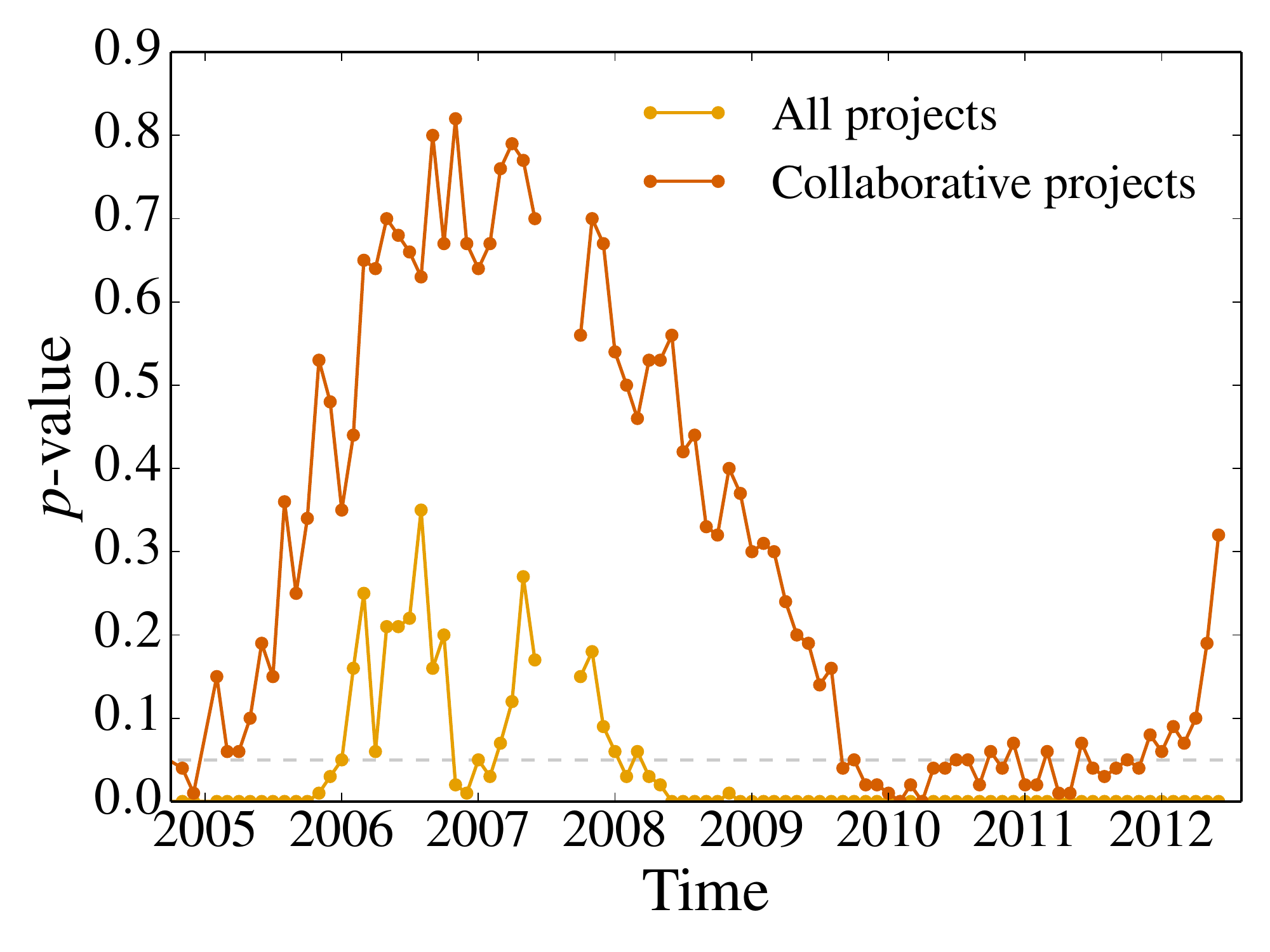}
  \caption{\label{fig:YulePval}
$p$-values for the goodness-of-fit test \citep{Clauset2009} of the Yule-Simon distribution and the empirical size distribution for each monthly snapshot. 
\emph{Collaborative projects} refers to the project size distribution with corrected value for the single-developer projects. 
We observe that the Yule-Simon distribution is a plausible candidate for the size distribution of all projects only for certain periods of time, while for the corrected empirical distribution it is a plausible candidate for most of the times, notably from the late 2005 to the late 2009.
}
\end{figure}

With respect to the \emph{project size distribution}, $f(x)$, we have to test the null hypothesis of the Yule-Simon distribution for each \emph{monthly} snapshot, an example of which is show in Fig. \ref{fig:CDF}(left). 
Simply fitting the Yule-Simon distribution to the empirical one and calculating the parameter $\rho_{\mathrm{all}}=3.88$  according to \citep{Garcia20118560} would lead to the result shown in Fig. \ref{fig:YuleFit} (lower right).
The (dashed-line) fit is visually worse, it does not capture the tail well because the (uncorrected) value of $f(1,N)$ is over-represented.

Therefore, in the next step we correct $f(1,N)$ as follows: 
Given the value $\rho_{\mathrm{all}}$ obtained from all projects, we first predict the value $f^{c^{\prime}}(1,N)$ for the collaborative single developer projects.
Then, we do a new fit of the  Yule-Simon distribution which leads to a corrected value $\rho_{\mathrm{col}}^{\prime}$. 
With this new value, we do a better prediction of $f^{c^{\prime\prime}}(1,N)$ and so forth.  
This method  is known as the \emph{expectation-maximization (EM) algorithm} \citep{Dempster1977},  where the number of single-developer projects $f^{c}(1,N)$ is used as an unknown, \emph{latent} variable.
EM is an iterative algorithm that consists of alternating expectation steps (E) and maximization (M) steps. 
Expectation refers to predicting $f^{c}(1,N)$, while maximization refers to calculating the appropriate $\rho_{\mathrm{col}}$. 
We halt the algorithm when the change of $\rho_{\mathrm{col}}$ is smaller then a given threshold  $\epsilon = 10^{-4}$.
This leads to the much better (solid gray line) fit shown in Fig.~\ref{fig:YuleFit} (lower right).
Taking all corrected distributions of Fig. \ref{fig:YuleFit} into account, 
we observe that the parameter $\rho_{\mathrm{col}}$ stays almost constant over time with 
values around $\rho_{\mathrm{col}}\approx 3$ (which contrasts with the uncorrected distributions where $\rho$ increases).  

Now, with the corrected size distribution $f^{c}(x,N)$, we can apply our rigorous goodness-of-fit test to each monthly snapshot. 
The results for the $p$-values are shown in Fig. \ref{fig:YulePval}. 
We see that the null-hypothesis of the Yule-Simon distribution as the  empirical one \emph{cannot} be rejected for all times between late 2005 and late 2009. 
This gives us great confidence both in our modeling assumptions and in the proposed correction to distinguish between collaborative and non-collaborative projects. 

At the same time, it leads us to the question what has changed after the end of 2009, to make the fits invalid. 
As we observe, from  2010 the significance level goes down considerably, although it is hardly really zero.
In order to better understand the dynamics from 2010, we will develop another conjecture in the following section.

\subsection{Estimations for  $p_0$}\label{sec:rhop0}

In the previous section we demonstrated that the Yule-Simon distribution (and the underlying model) is a valid candidate for describing the empirical dynamics of collaborative projects at least for certain time intervals. 
We can link our findings for the size distribution, which refer to the \emph{systemic} level, back to our assumptions for the \emph{microscopic} dynamics. 
Recall that in the model of Simon \citep{simon1955} there is only one parameter $p_{0}$ that decides whether new developers found new projects, as opposed to joining existing ones. 
This parameter, as far as the theory goes, is directly linked to the exponent $\rho$ of the distribution, via Eq. (\ref{rho}). 
Assuming $\rho=3$, Eq. (\ref{rho}) would give $p_{0}=2/3$, which is quite high if compared, e.g., to the \emph{firm size distribution} where $\rho$ is about 1.2 and $p_{0}$ about 0.16.

In this section we want to find an independent way of estimating $p_{0}$.
We recall that $p_{0}$ essentially describes how much of the total growth goes into newly founded projects. 
So, if $G_{\mathrm{tot}}$ is the growth spent on all existing projects and $G_{1}$ is the growth spent on new projects during a given time interval, then $p_{0}={G_{1}}/{G_{\mathrm{tot}}}$ \citep{simon_bonini_1958}.

In our empirical data, $G_{\mathrm{tot}}(t)$ is measured by the total number of \emph{developers} per month that enter SF, $\Delta N_{d}(t)$ which is shown in Fig. \ref{fig:Project EnEX Interval=1}(right). 
$G_{1}(t)$, on the other hand, is given by the total number of newly founded projects per month, $\Delta N_{p}(t)$, shown in Fig.~\ref{fig:Project EnEX Interval=1}(left). 
So, we just divide these monthly entry rates, to obtain $p_{0}(t)=\Delta N_{p}(t)/\Delta N_{d}(t)$ from the empirical data. 
We do this for the two different datasets: (a) for all \emph{all} projects, and (b) for \emph{collaborative} projects.
For the latter, we need to correct also $\Delta N_{d}(t)$ because we have to \emph{exclude} those developers that joined SF to establish a non-collaborative project. 
These correction is done based on the empirical data.

\begin{figure}
  \centering
  \includegraphics[page=3,width=0.49\textwidth]{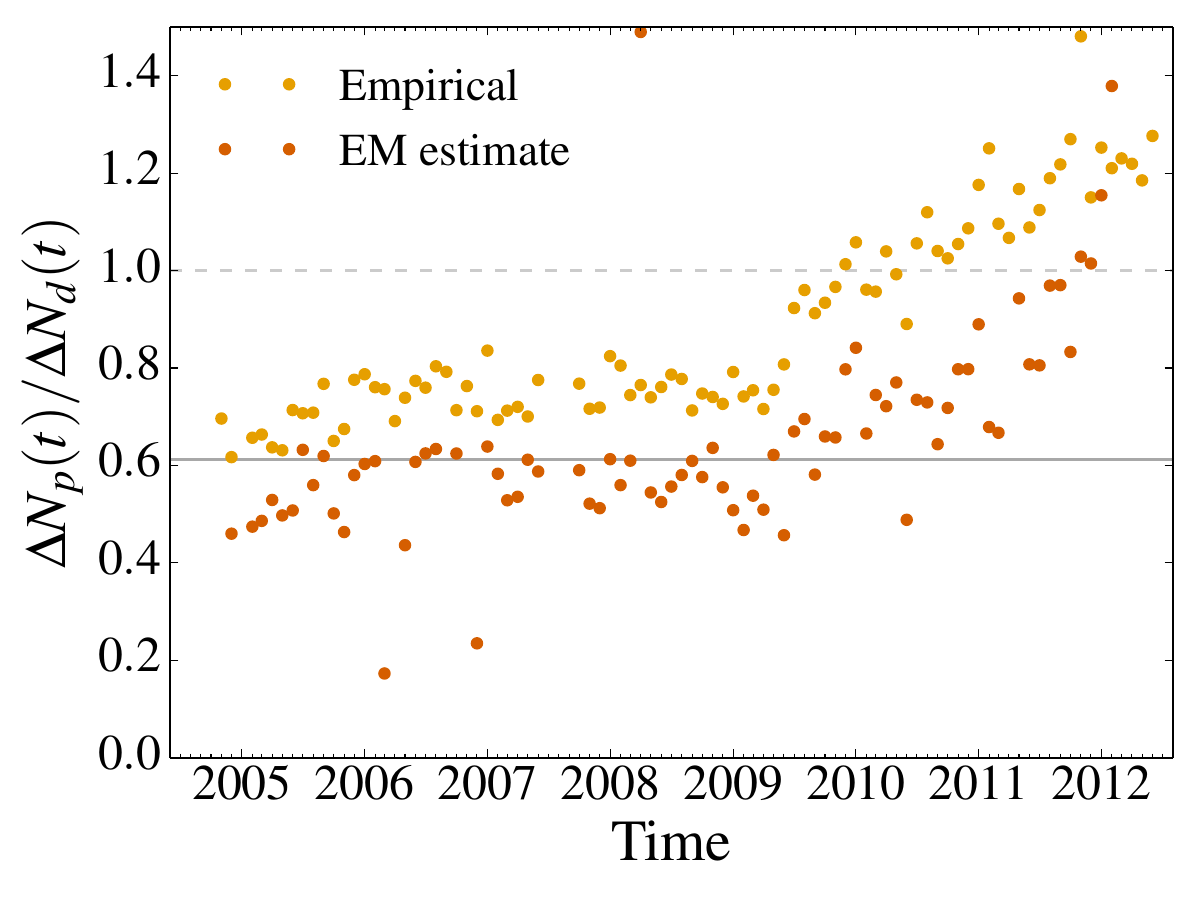}

  \caption{\label{fig:YuleTMP}
Estimation of the parameter $p_{0}$ from the monthly entry rates of projects and developers, both for all projects/developers ( $p_{0}^{a}(t)$, yellow dots) and for collaborative projects/developers only ( $p_{0}^{c}(t)$, red dots).
The full line is the median of $p_{0}^{c}(t)$ at {0.6128}.}
\end{figure}

The results are plotted in Fig. \ref{fig:YuleTMP} for the two different data sets: (a) $p_{0}^{a}(t)$ for all \emph{all} projects and developers, and (b) $p_{0}^{c}(t)$ for the collaborative  projects and developers. 
As expected from the above discussion, we see differences for the time periods before and after 2010. 
In fact, we see that after 2010  $p_{0}^{a}(t)$ has consistently values \emph{above} 1, which \emph{cannot} be realized from the assumption that only newly entering developers establish new projects. 
If the latter holds, $p_{0}^{a}$ is necessarily bound to values below 1.
To explain this, we arrive at our second conjecture that after 2010 an increasing number of \emph{established} developers started to found \emph{new} projects.
This, however, is not considered in our modeling assumptions, therefore the prediction derived from the model necessarily fails as we also see from the low $p$-values in Fig. \ref{fig:YulePval}.

If we look at \emph{collaborative} projects, we see that  $p_{0}^{c}(t)$ follows the same trend as  $p_{0}^{a}(t)$, just with a shift toward lower values. 
To find out whether this shift results from a higher entry rate of collaborative projects or a lower entry rate of collaborative developers, we have plotted in Fig. \ref{fig:OSS Growth Vs. Time}(right) $\Delta N_{p}(t)$. 
We verify that $\Delta N_{p}(t)$ for collaborative projects is almost constant over all years, i.e., the increase in  $p_{0}^{c}(t)$ is from the lower entry rate of collaborative developers.  
Because of the similar trend compared to $p_{0}^{a}(t)$ after 2010 we also keep our conjecture that an increasing number of established developers started to found new \emph{collaborative} projects. 
This violation of our modeling assumption can be confirmed also in Fig. \ref{fig:YulePval}, where we see that the $p$-values for the goodness-of-fit test break down after end of 2009 for the collaborative projects. 
If we take the median for  $p_{0}^{c}(t)$ over the whole time period, we find $\bar{p}^{c}_{0} \approx 0.61$ which is in a remarkable agreement with the theoretical value $p_{0}=2/3$ obtained from $\rho=3$.

Eventually, we want to discuss an additional issue in comparing the empirical and theoretical results. 
By means of the EM method, we found a way to correct $f(1,N)$ such that only collaborative projects are taken into account. 
The corrected value $f^{c}(1,N)$ can also be related to the empirical number of collaborative single-developer projects. 
I.e., by tracing their history, we can identify in the data set those single-developer projects that grew at a later point in time. 
Their monthly growth rate $\Delta N_{p}$ is already plotted in Fig. \ref{fig:OSS Growth Vs. Time}(right) (blue line). 
We observe that there is a shift between the predicted growth rate of collaborative single-developer projects (green line) and the empirical one, which slightly increases over time from values of 1.5 to 2.

The cause for this mismatch should not be attributed to the predicted value, but rather to the empirical one because it  \emph{underestimates}  the number of collaborative projects for the following reason. Our empirical classification of collaborative vs. non-collaborative projects is based on their observed growth, only. 
If we classify projects as non-collaborative, we make a mistake because projects may still grow later in time, but this is just not observed.
This mistake becomes larger the closer we come to the end of the data set. 
Therefore, to estimate the magnitude of the mistake, we should look at the oldest projects in the data set, which date back to November/December 2004 (when the data became reliable). 
We verify that the interval between the time when these projects were created and the time when a second developer joined, can be well described by an exponential distribution, $f(t) = \lambda 
\exp^{-\lambda t}$. 
The expected value of this distribution is $\mathbb{E}[t] = {1}/{\lambda}$, which can be also measured from the data for the almost 3000 projects created in these two months, to yield roughly 450 days. 
The cumulative distribution gives us the probability that those projects will grow before the 450 days as $\Pr(t < \mathbb{E}[t]) \approx 0.6$.

This can now be used to calculate the mistake made by classifying projects as non-collaborative during the last 450 days of the data set. 
It is about $40\%$, i.e., we miss 40\% of the collaborative projects in our estimation. 
The correction for the observed number thus will be a factor of $1/0.6 \approx 1.6$, which is very close to the observed mismatch of 1.5 to 2.
I.e., with this rough estimation we can explain fairly well the observed difference between the empirical and EM estimates.

To conclude our discussion so far, we find that the Yule-Simon distribution is a valid candidate to describe the empirical size distribution of \emph{collaborative} projects. 
However, this validity is constrained to certain time periods for which the underlying assumptions of the model can be justified. 
We observe that in later time periods established developers started to create additional projects.
This was not considered in the model assumptions to derive the Yule-Simon distribution, where only newly entering developers are considered to create new projects.

\subsection{Extensions}
\label{sec:ext}

In this paper we have investigated to what extend an established model for the entry of new \emph{firms} and the growth of existing ones originally developed by Simon \citep{simon1955} can be directly applied to OSS communities, where new projects are founded by new developers and existing projects grow by attracting new developers.

The advantages of using the Simon model go alongside with the disadvantages resulting from the limitations arising from the underlying assumptions. 
We discuss them here, to give hints for further improvements of the model, because we noticed that the Yule-Simon distribution, over large time intervals, has shown to be a promising candidate for the size distribution of collaborative projects. 
However, each of the suggestions discussed below will modify the original model such that the analytical approach developed can no longer be used and closed-form solutions will most likely not be derived.  

The \emph{first} suggestion relates to a known criticism of the Yule-Simon model, namely that not more than one project can be founded or grow at each time step. 
This does not make problems as long as one is interested in the asymptotic size distribution.
But in order to come up with a more realistic dynamics before the steady state is reached, one should consider that projects can be founded and grow in parallel. 
In particular, one has to consider concurrent activities, i.e., that not only new developers perform an action, but also established developers can decide to have more than one project, e.g., by founding a new one. 
As a consequence, $p_{0}$, the probability to found a new project, shall become a \emph{heterogeneous} parameter, to better reflect individual motivations of developers.
By this, we could further distinguish between different personalities, e.g., \emph{founders}, who prefer to start new projects, and \emph{contributors}, who prefer to join existing projects. 

As a \emph{second} suggestion, we can consider that new developers may have a \emph{preference} for larger or smaller projects, i.e., the probability to choose from size class $(x-1)$ is also proportional to $(x-1)^{\alpha}$.
This was already implemented in the dynamic assumptions of Eq. (\ref{eq:pn1}) as a new element not discussed in \citep{simon1955}.
$\alpha=0$ would recover the case of \emph{no size preference} (as, e.g., also used to describe the firm growth dynamics \citep{sutton}), $\alpha=1$ would be a preference directly proportional to the existing size, and $\alpha<0$ would indicate that projects become \emph{less attractive} with increasing size, probably because of coordination and integration efforts.
Hence, the additional parameter $\alpha$ allows us to consider various (monotonous) \emph{size-dependent preferences}.
To account for optimal project sizes, this dependence should be also non-monotonous. 
In our formal approach, we have set $\alpha=1$ to favor the mathematical approach by which the project size distribution was derived. 
Without this restriction to closed-form solutions, an agent-based simulation could explore the impact of size-dependent preferences on the project size distribution. 

As a \emph{third} suggestion, we should allow contributors to \emph{switch} between projects, in order to better utilize their skills.
This would also have consequences for the \emph{knowledge spillovers} between projects, which is an important consideration for management science and economics. 
On the formal modeling level, such additional assumptions would change the rate equation approach developed in Sect. \ref{sec:rate}, by adding additional terms for the shrinkage of existing projects (developers leave) and for the growth of existing projects by other than newly registered developers. 

A different set of possible extensions points to the way the developer \emph{activity} is counted in. 
So far, we have assumed that a newly entered developer \emph{immediately} founds a new project or joins an existing one. 
But developers may have joined SF for different other reasons, e.g., for getting better access to code to re-use outside of SF. 
This may lead to a mismatch between the number of developers entering SF per month and the assumed activity of these developers inside SF. 
In the same line, in our analysis developers are assigned to projects, which is indicated by a link, and our modeling approach assumes that such links do not change. 
However, links do not necessarily mean that developers are actively working for the project, they are only a first (and not necessarily the best) approximation of contributions. 
Here, we note that already more refined measures are available which are discussed in a subsequent paper \citep{github}.
But these measures largely depend on information that is not available from \texttt{Sourceforge.net}, so we will have to resort to \texttt{Github.com}

We conclude that, even with these limitations on the SF data, our analysis about the launch of new projects and their subsequent growth is one of the largest investigations on such data to date. 
It resulted in new findings about the project size distribution and the degree distribution of developers, about their entry and exit rates, the preferred usage of programming languages.  
In order to better understand the dynamics that generated these systemic properties, we utilized an established economic model that in this paper has proven to be a valuable candidate also for the modeling of  socio-technical systems such as OSS communities.
At the same time, the model also revealed some shortcomings which helps us to better understand the role of underlying assumptions and their limitations. 
At the end, not only the (positive) confirmation, but also the (negative) rejection of modeling assumptions both generate important insights into the dynamics of real socio-technical systems.

\subsection*{Acknowledgements}
\label{sec:Acknowledgements}
Early investigations on this topic were supported by the Swiss National Science Foundation (Grant CR12I1\_125298).
We thank Natalia Frey for discussions about the Simon model.

\bibliographystyle{ws-acs} \bibliography{items}

\end{document}